\documentclass[aip,reprint]{revtex4-2}

\usepackage[dvipdfmx]{graphicx}
\usepackage{bmpsize}
\usepackage{color}

\usepackage[utf8]{inputenc}
\usepackage[T1]{fontenc}

\begin{document}

%\preprint{}

\title{AI-equipped scanning probe microscopy for autonomous site-specific atomic-level characterization at room temperature} %Title of paper



\author{Zhuo Diao}
\email[]{enzian0515@gmail.com}
%\thanks{}
\affiliation{Graduate School of Engineering Science, Osaka University, 1-3 Machikaneyama, Toyonaka, Osaka 560-0043, Japan}

\author{Keiichi Ueda}
%\thanks{}
\affiliation{Tokyo Metropolitan Industrial Technology, Research Institute, 2-4-10 Aomi, Koto-Ku, Tokyo, 135-0064, Japan}

\author{Linfeng Hou}
\author{Fengxuan Li}
\author{Hayato Yamashita}
\author{Masayuki Abe}
\email[]{abe.masayuki.es@osaka-u.ac.jp}
%\thanks{}
\affiliation{Graduate School of Engineering Science, Osaka University, 1-3 Machikaneyama, Toyonaka, Osaka 560-0043, Japan}


%\date{\today}

\newpage

\begin{abstract}
We present an advanced scanning probe microscopy system enhanced with artificial intelligence (AI-SPM) designed for self-driving atomic-scale measurements. This system expertly identifies and manipulates atomic positions with high precision, autonomously performing tasks such as spectroscopic data acquisition and atomic adjustment. An outstanding feature of AI-SPM is its ability to detect and adapt to surface defects, targeting or avoiding them as necessary. It's also engineered to address typical challenges such as positional drift and tip apex atomic variations due to the thermal effect, ensuring accurate, site-specific surface analyses. Our tests under the demanding conditions of room temperature have demonstrated the robustness of the system, successfully navigating thermal drift and tip fluctuations. During these tests on the Si(111)-(7$\times$7) surface, AI-SPM autonomously identified defect-free regions and performed a large number of current-voltage spectroscopy measurements at different adatom sites, while autonomously compensating for thermal drift and monitoring probe health. These experiments produce extensive data sets that are critical for reliable materials characterization and demonstrate the potential of AI-SPM to significantly improve data acquisition. The integration of AI into SPM technologies represents a step toward more effective, precise and reliable atomic-level surface analysis, revolutionizing materials characterization methods.
\end{abstract}

\keywords{self-driving, deep learning, scanning probe microscopy, scanning tunneling spectroscopy, room temperature}

\maketitle %\maketitle must follow title, authors, abstract

\section*{Introduction}

The integration of artificial intelligence (AI) with nanotechnology has been recognized as crucial since the pioneering work by Drexler in 1986\cite{Drexler1986Engines}. However, the advancements described in this book had not been widely realized.
As a key tool of the nanotechnology, scanning probe microscopy (SPM) has emerged in characterizing nanoscale surfaces, enabling the discovery of new surface properties and phenomena\cite{stm, afm}. 
Today, SPM has significantly contributed to both basic science and industrial applications. 
Especially, in the field of basic science, it can not only image surfaces, but also measure the physical properties of individual atoms and move atoms to create structures\cite{Eigler1990}. Most of these experiments have been conducted in cryogenic environments. This is because not only the SPM equipment but also the tip of the probe and the sample itself are thermally stable.
However, from a practical standpoint, conducting these processes at room temperature is essential.

Even in room temperature environments, where thermal effects can affect measurements, SPM has provided significant capabilities such as dynamic imaging to observe temporal changes in chemical reactions\cite{chem1, chem2}, biological processes\cite{hsafm}, surface dynamics\cite{dynamics1, dynamics2}, diffusion\cite{BrihuegaGeDiffusion, SnOnSi}, and crystal growth\cite{surface_growth}.
In addition, there are studies of dopant atom manipulation that can be done at room temperature\cite{am1, am2}.

There have been challenges in further pursuing these pioneering room-temperature experiments.
Using the room-temperature SPM to achieve atomic resolution does not remove the thermal effects on the device, which lead to measurement instability.
One of the most significant effects of thermal fluctuations in atomic-resolution SPM measurements is thermal drift. 
Thermal drift continuously changes the relative position of the tip and sample atoms, which not only prevents continuous measurement of the same area, but can also cause image distortion. 
Another problem of the room-temperature SPM is frequent change of the tip apex. 
The quality of the image is affected by the frequently change of the tip apex atom especially in atomic resolution imaging.
Repairing the tip apex is usually performed by touching to the surface, which requires time and a great deal of attention.
To achieve the high-precision measurements under such inherently non-optimized conditions, SPM techniques such as drift correction\cite{drift1, drift2} and tip fabrication\cite{sharp1, sharp2} become indispensable in the context of room temperature SPM. 
However, even with these techniques, it was very difficult to perform site-specific experiments at the atomic level at room temperature.

%For example, scanning tunneling spectroscopy (STS), a technique for site-specific measurements using SPM, enables the assessment of local electronic states. Typically, STS is performed at very low temperatures where the thermal instabilities of both the STM system and the sample under examination become negligible.
%On the other hand, differences between low-temperature and room-temperature STS data have been reported for a Si(111)-(7$\times$7)  surface\cite{PhysRevB.73.161302}. Discussing material properties necessitates an understanding of the local electronic structure at room temperature, particularly concerning the effects of dopants and defects. Moreover, even when site-specific data are obtained under optimal conditions, the results may still reflect the impact of thermal fluctuations. Therefore, acquiring a substantial volume of data and subjecting it to statistical analysis is crucial for generating reliable results. However, in our current SPM system, gathering extensive amounts of site-specific data at room temperature proves to be quite challenging.

In recent years, to overcome the limits of human capability, experiments utilizing the concept of the self-driving laboratory\cite{selfdriving1, selfdriving2} have become increasingly prevalent, driven by the demand for extensive and complex data collection. This approach has risen to prominence as a key solution for challenges like new materials discovery, largely through the application of artificial intelligence (AI). 
The self-driving concept is also anticipated to be highly beneficial for time-consuming and labor-intensive SPM experiments, which is  manually operated.
A deep learning model, adept in signal processing and computer vision, can identify specific patterns with accuracy nearly on par with human experts\cite{resnet, vgg}. Strategically integrating AI can significantly reduce the dependency on manual intervention in SPM operations\cite{spm_PERSPECTIVE}.

There are previous studies of the integrating the image recognition AI and SPM yielding notable contributions to the domains of data analysis and processing\cite{ai_ana1, ai_ana2, ai_ana3, ai_ana4, ai_ana5}.  The application of these AI-driven analysis techniques in the processing of data measured in real-time holds the potential to significantly automate SPM operations. The efficacious deployment of such technologies requires intricate integration within the control mechanisms, encompassing both hardware and software, that includes scanning protocols and AI features, alongside the acquisition of extensive datasets for the cultivation of high-efficiency AI models. Particularly in cryogenic conditions, where thermal disturbances are markedly reduced, recent research has underscored the utility of AI-facilitated methods in promoting autonomous scanning procedures \cite{ai_sts, ai_scan, ai_scan2} and the manipulation of individual atoms\cite{ai_am}.
However, unlike setups in cryogenic environments, more sophisticated AI-based SPM (AI-SPM) is required in room temperature environments. This enhancement is necessary to compensate for challenges posed by thermal fluctuations and to ensure the reproducibility of experiments.
Robustness in handling measurements under unstable conditions is crucial for performing real-time and site-specific experiments at room temperature. Moreover, it is essential to recognize during dataset collection that the simulated datasets used for training\cite{ai_ana2, ai_ana4}  may not fully capture the variations in atomic images. This limitation arises from differences in the tip state and the patterns of adsorbates in actual measurements. Addressing these considerations is vital for enhancing the reliability and performance of site-specific SPM measurements at room temperature.

In this manuscript, we present a deep learning-assisted AI-SPM system, specifically designed for the site-specific operation at room temperature. Our AI-SPM system is an integrated fusion of software, control firmware, and hardware components, which facilitates the development of a robust neural network through a systematic dataset collection process. 
Decision-making and optimization tailored for room temperature conditions, is fully automated, with the objective of achieving atomic precision measurements.
Utilizing this system, we demonstrate two key applications in surface characterization: autonomous acquisition of high quality images and large data sets for atomic precision scanning tunneling spectroscopy (STS) at room temperature.

\section*{AI-SPM for self-driving measurement}

\begin{figure}[ht]
\centering
\includegraphics[width=100mm]{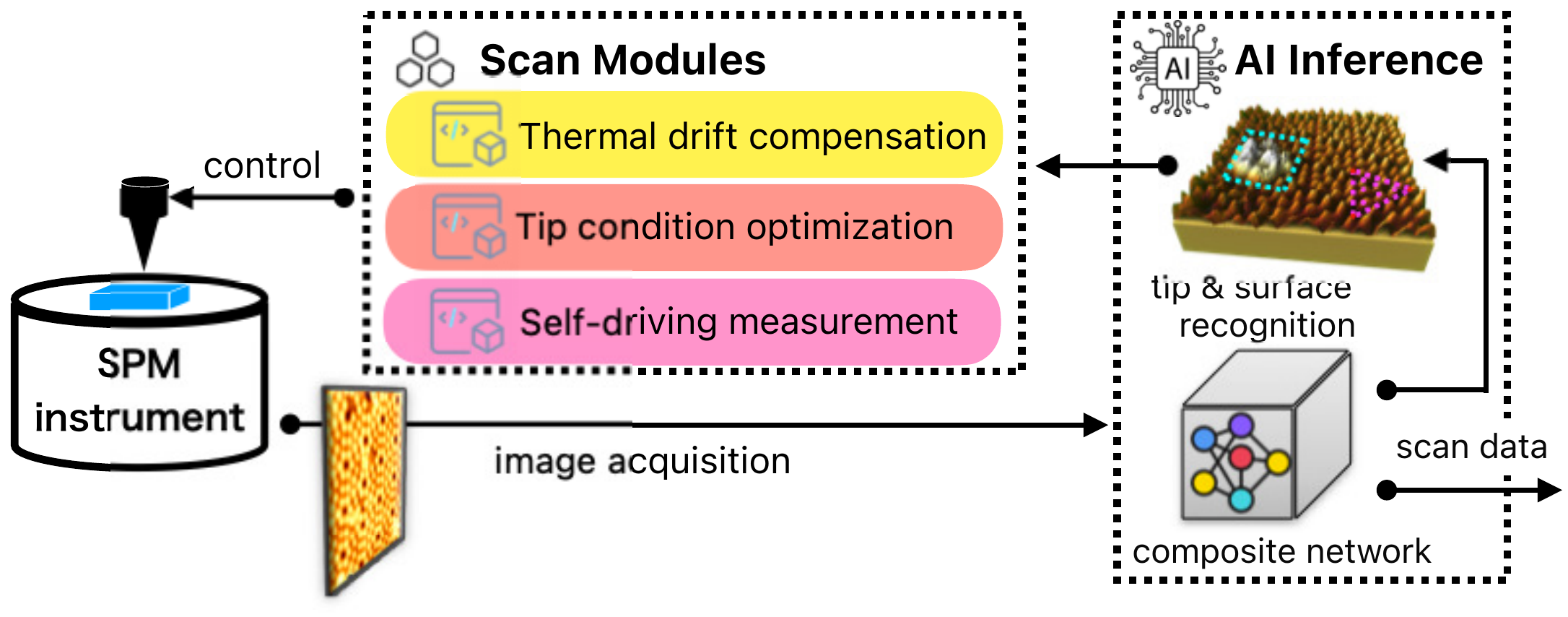}
\caption{Schematic flow of artificial intelligence scanning probe microscopy (AI-SPM) system.
The system compensates for thermal drift and the tip apex, which is a problem inherent to room temperature. Then, It also determines the presence of adsorption sites and defects, and performs site-specific experiments at targeted atomic positions. Large amounts of data can be obtained automatically.}
\label{fig:aispm}
\end{figure}

\subsection*{AI-SPM configuration}

Figure~\ref{fig:aispm} illustrates the configuration of our AI-SPM system.
SPM hardware is conventional one, and two main program parts are added to our home-built scan software for the AI operation.
The AI inference section receives the SPM measurement data, makes a situational judgment and determines the next task. In particular, it uses the images to determine the state of the probe tip, the identification of individual atomic sites and unit cells, the location and type of adsorbates, and whether further site-specific measurements are possible.
The information of the tip and sample surface sent to the Scan Module part from the AI inference component is utilized in the experiment programmed by the operator. In Fig. \ref{fig:aispm}, the Scan Module comprises two scripts essential for the room temperature experiment (thermal drift compensation and tip condition optimization) alongside a self-driving measurement script.
These scripts are used in the experiments in this manuscript.

\subsection*{Convolutional neural networks on AI inference}

The acquired image data of the surface not only confirms the crystal structure of the surface, but also provides information of the condition for the site-specific measurement: presence of defects or adsorbates, not atomically clean area or steps, tip apex condition.
To automatically make decisions on all this information, we have employed convolutional neural networks (CNNs). Each CNN is tailored for a specific predictive task, and they all utilize scanned topography as their input. By integrating these CNNs into a composite network, we enable comprehensive access to a wide range of information about the real-time scanning topography\cite{tipfix}. Understanding of both tip and surface conditions enables the site-specific measurements such as STS and atomic manipulation.

\begin{figure}[h]
\centering
\includegraphics[width=0.9\linewidth]{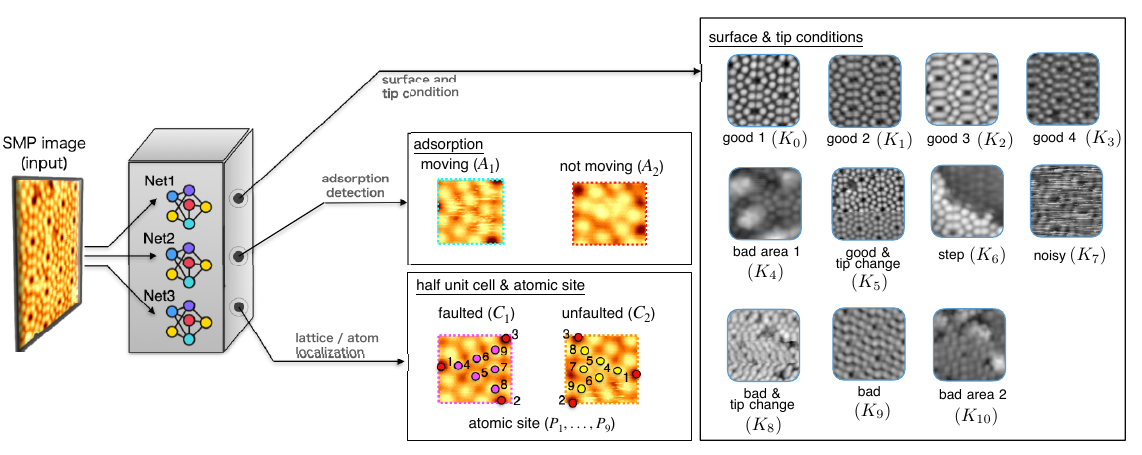}
\caption{
Convolutional neural networks (CNNs Net1, Net2, and Net3) employed in this study for room-temperature site-specific analysis of the Si(111)-(7×7) surface. Net1 is tasked with recognizing the conditions of the tip and sample, Net2 identifies adsorption sites, and Net3 is responsible for recognizing half unit cells and atomic positions.
}
\label{fig:model_architecture}
\end{figure}

In Fig. \ref{fig:model_architecture}, the CNN architecture for the Si(111)-(7$\times$7) surface is presented. We designed a composite network structure comprising three distinct models—Net1, Net2, and Net3—each dedicated to a specific task.
Net1 is tasked with recognizing the conditions of both the tip and sample, performing a multi-class classification of tip and surface conditions $K_i (i=0, 1,\cdots, 10)$\cite{tipfix}.
It determines whether the sample and the probe are both in a state ($K=1, 2, 3, 4$) in which site-specific measurement is possible. If the surface is not contaminated and the tip is capable of atomic resolution, it is considered to be in  a ``good" condition for site-specific measurement.
Our previous studies have demonstrated that the accuracy of Net1 is from 87\% to 90\%.
The reason for classifying good into four categories is to determine and create a probe that can be used for atomic manipulation in the future\cite{doi:10.1021/nn403097p, PhysRevB.78.205305}.

Net2 determines adsorbates. On the Si(111)-(7$\times$7), we can find some adsorbates at the atomic level that are moving and some are not. 
Here, we define the one imaged brighter than the Si adatom in the half-unit cell to be the immobile adsorbent $A_2$.
As another type of adsorbate, there are ones that are imaged like noise or feedback error. These are adsorbates in motion remaining in a half-unit cell ($A_1$ in Fig. \ref{fig:model_architecture}). 
The moving adosorbates in the half unit cell have been studied for experiments with clusters and atomic manipulation\cite{doi:10.1021/nn202636g, sugimoto2014NatCom, Inami2015NatCom, Hwang1999DYNAMICBO, PhysRevLett.94.176104, PhysRevLett.95.146101, PhysRevLett.101.266107, Osiecki2022NatCom, CUSTANCE20011406}.
Net2 is designed to be applied to these studies as well.

Net3 detects site-specific information of the Si(111)-(7$\times$7) surface.
Half-unit cells are categorized as bounding boxes of $C_1$ and $C_2$, respectively.
Applying a negative sample bias during measurement and observing the screen's response allows for the determination of whether $C_1$ or $C_2$ corresponds to the faulty or non-faulty section. In the experiment of Fig. \ref{fig:model_architecture}, $C_1$ was identified as faulted, while $C_2$ was found to be unfaulted.
Local atomic sites are represented as key points  $P_i\ (i=1,2, \cdots, 9)$(points of coordinate $(x, y)$): three corner holes around each half-unit cell as $P_1, P_2, P_3$, corner adatom as $P_4, P_8, P_9$, and center adatom as $P_5, P_6, P_7$.

\subsection*{Scan Module Scripts}

The Scan Module comprises two scripts essential for the room temperature experiment (thermal drift compensation and tip condition optimization) alongside a self-driving measurement script.
These scripts are used in the experiments in this manuscript.
Users can add own script to the Scan Module to customize the experiment. 

The module for the thermal drift compensation works based on the feedforward technique present in our previous study\cite{drift}.
Continuously obtained SPM images are compared using the feature point matching algorithm to output the thermal drift velocity at the minute scale, enabling the correction of even non-linear thermal drift at the days scale.

It is equipped to autonomously maintain the optimum state of the probe\cite{tipfix}. Here, the CNN determines the state of the tip from the images obtained: if the CNN determines that the tip is not optimal for atomic resolution measurements, a tip shaping is performed by bringing the tip close to the surface and simultaneously changing the bias current. The images are then acquired and the Net1 judges the state of the tip. These processes can be performed automatically until the tip becomes a ``good" condition.
The thermal drift module is activated during the tip optimization process for automatically compensating for thermal drift.

The self-driving measurement module autonomously performs data acquisition based on the AI Inference output and can also leverage other module functionalities for drift correction and tip condition optimization. 
Therefore, our AI-SPM system represents a departure from traditional automation methods with fixed routines. It offers the ability for self-guided data exploration and acquisition, adapting to previous scans through a closed-loop mechanism. Detailed descriptions of the AI-SPM hardware are provided in the Methods section.

\subsection*{Two-phase training data acquisition}

As mentioned above, at room temperature, even if thermal drift is compensated, we are facing unstable image conditions due to tip apex change.  Adsorbates and defects may be present in the first place, but they may also appear during scanning, significantly influencing the SPM images. This makes us assemble a comprehensive and varied set of training data is needed.
To acquire data that take into account both the number and variety of datasets, 
we have employed developed ``two-pase" approach to training the AI-SPM itself as the experiment progresses.

Phase 1 comprises collecting and creating a dataset comprising image information obtained under diverse conditions. 
Here, every image is acquired, including those that have not been atomically resolved.
To obtain a lot of images categorized $K_0$ to $K_{10}$, same process used in the tip-condition optimization was conducted.
In this phase,  although the initial accuracy may not fully optimize the tip and sample state, this process continues iteratively, accumulating an extensive dataset. 
This dataset is essential in training a robust model that can distinguish data quality, ensuring the system's capability to autonomously capture atomic-resolution images on Si(111)-(7$\times$7).

The second phase is intended to further enhance the data set of images being obtained at atomic resolution. 
Autonomous measurement on a target atom is performed within the self-driving measurement module, with the assistance of Net3, which provides information about atom locations. 
One of the examples of measurement is
autonomous atom manipulation techniques. 
It utilizes the "atomic pen" technique\cite{am2} to manipulate Si atoms within a unit cell and individually place in-motion Si atoms as single adsorptions.
By repeatedly executing phase 2, a larger number of adsorbate and defect patterns on the topography are generated, augmenting the datasets utilized in training Net1, Net2, and Net3.

To enhance the accuracy of Net1 and Net3, the dataset has been expanded by incorporating topography scans obtained from various leading states and scan areas. In phase 1 and phase 2, 11,616 and 7,269 images were acquired, respectively. From these, 2,082, 255, and 545 images and their augmented data were used in the training of Net1, Net2, and Net3, respectively.
These data points, likely consisting of individual images or instances, comprehensively cover a wide range of surface properties.
The performance evaluation using confusion matrices of Net1, Net2, and Net3 is detailed in the supplementary information (see Fig. \ref{fig:model_performance} in Supplementary \ref{sup:model}), demonstrating that they achieved high scores.

\section*{Implementation of AI-SPM}

\subsection*{Local site identification}

\begin{figure}[h]
\centering
\includegraphics[width=0.7\linewidth]{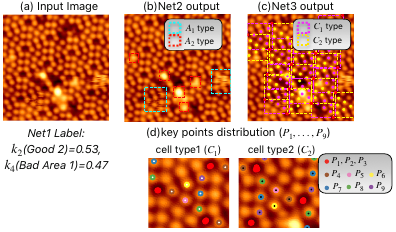}
\caption{
Performance of the local site identification on $\mathrm{Si(111)-(7\times 7)}$ surfaces.
(a) an STM image after Net1 is evaluated, which is used for the site identification as shown in (b) to (d). 
The evaluation results are shown below the image.
(b) Identification of adsorbates by Net2. $A_1$ and $A_2$ indicate moving and stationary adsorbates, respectively. 
(c) Net3 classification of half-unit cell type.
$C_1$ and $C_2$ are faulted and unfaulted half unit cells, respectively. 
(d) Site-specific identification of adatom site and corner holes in the half-unit cells.
Each site is labeled by Net3 as $P_i (i = 1 \cdots 9)$.
Sample bias voltage and tunneling current in the STM measurement are $V_s$=1.5~V and  $I_t$=200~pA, respectively.
}
\label{fig:model_output}
\end{figure}

Performance of the local site identification of STM measurement at room temperature on $\mathrm{Si(111)-(7\times 7)}$ surfaces is shown in Fig. \ref{fig:model_output}. 
Figure \ref{fig:model_output} shows the capability of the trained Net1, Net2, and Net3 that recognize key points of the surface for  the site-specific measurements
Images acquired as shown in Fig. \ref{fig:model_output} (a) are evaluated by Net1. In this image, the  $\mathrm{(7\times 7)}$ structure and defects and adsorbates are present. For this image, Net1 outputs weight values of $k_2 \mathrm{(Good2)} = 0.53$ and $k_4 \mathrm{(Bad\ area)} = 0.47$. This means that the tip condition is good, but surface defects and adsorbates are present.
In identifying adsorbates in Net2, it can be seen in Fig. \ref{fig:model_output} (b) that it can identify stationary adsorbates ($A_1$), surrounded by red dashed lines, and adsorbates diffusing in the half-unit cell ($A_2$), surrounded by light blue dashed lines.
In Fig. \ref{fig:model_output} (c), Net3 classifies almost all regions of the image into $C_1$ (faulted half) and $C_2$ (unfaulted half) categories, as well as identifying individual adatoms and corner holes within these halves.
Net3 is capable of identifying each adatom. As depicted in Fig. \ref{fig:model_output} (d), the individual adatoms and corner holes are represented by $P_i (i=1,2,\cdots,9)$. The labels indicate that $i=1, 2, 3$ correspond to corner holes, $i=4, 8, 9$ to corner adatoms, and $i=5, 6, 7$ to center adatoms.
These results mean that our proposed Net1, Net2, and Net3 methods can identify individual atomic sites as keypoints at room temperature.

\subsection*{Probing for optimal measurement regions}

\begin{figure}[h]
\centering
\includegraphics[width=0.8\linewidth]{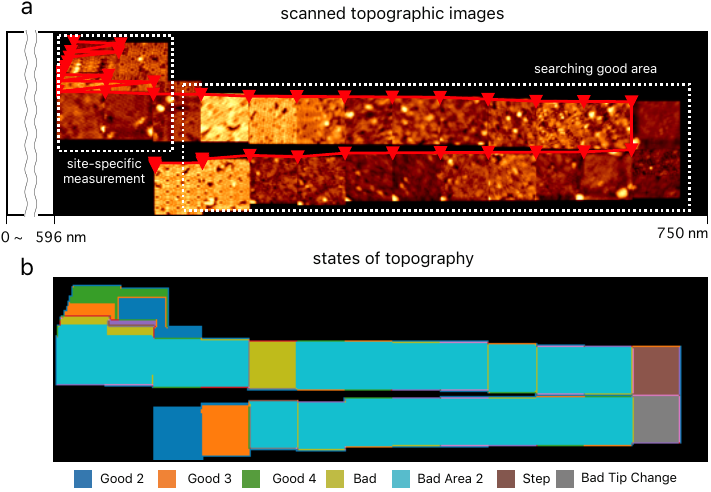}
\caption{
Sequential images of STM on the $\mathrm{Si(111)-(7\times 7)}$ surface being measured while automatically searching for the optimum measurement area and tip conditions.
(a) Sequential images measured by self-guided STM.
The first scan starts at the upper right corner, and the scan area is automatically changed.
The upper left portion of each image is indicated by an inverted triangle, and the inverted triangles are connected by red lines to show the trajectory. 
(b) Top-most inferred state during the whole data acquisition process including tip conditioning, drift compensation, and STS target atom localization during continuous scan expressed in color.
All STM topographic images were acquired with 2~V sample bias, -200~pA set point, 105~s scan time, and 11.25 $\times$ 11.25~nm scan area.
During this measurement, thermal drift correction is activated by feature point matching\cite{drift}.
}
\label{fig:demo1_auto_scan}
\end{figure}

Figure \ref{fig:demo1_auto_scan} shows a continuous image of the room temperature STM measurement on the $\mathrm{Si(111)-(7\times 7)}$ surface, in which atomic images free of defects, adsorbates, and steps are autonomously identified.
In the experiment, a total of 45 consecutive images were acquired over different regions. 
In Fig. \ref{fig:demo1_auto_scan}(a), scanning started from the upper left corner of the figure, with the upper left portion of each image marked by an inverted triangle. The trajectory of the regions scanned is shown by red lines connecting the sequence of images.
Here, the scanning routine includes two critical modules to identify the atomic resolution images. First, a thermal drift correction module compensates thermal drift to minimizes the image distortion\cite{drift}. 
As the second module, Net1 keeps the tip in a state conducive to atomic resolution measurements while bypassing areas affected by impurities, atomic defects, and step edges. 
Figure \ref{fig:demo1_auto_scan}(b) shows Net1 judgment of the acquired STM images, classifying the most plausible probe or surface conditions shown in Fig. \ref{fig:model_architecture} by color.

Until now, it has been necessary to determine the appropriate measurement area and to continuously monitor the state of the probe tip. However, unlike in cryogenic environments where the probe tip remains relatively stable, at room temperature the probe tip frequently changes. This makes it virtually impossible for researchers to conduct sustained long-time experiments.
As demonstrated in Fig. \ref{fig:demo1_auto_scan}, by integrating advanced automation and deep learning techniques, we have been able to establish optimal experimental conditions for exploring regions with atomic resolution.
This method enables the automatic identification of the appropriate probe and region at room temperature, which is essential for conducting site-specific experiments. Consequently, experiments such as atomic manipulation and STS are expected to be feasible under conditions comparable to those in cryogenic environments.

\subsection*{Self-driving scanning tunneling spectroscopy at room temperature on different Si(111) adatom sites}

Using the methods described so far, the basic tools are now available to realize an AI-SPM capable of self-steering measurements.
As an example of a site-specific measurement at room temperature, we perform $I-V$ measurements on adatoms of the Si(111)-(7$\times$7). 
From the obtained $I-V$ curve, the data of dI/dV, or STS as the local electronic state, can be calculated.There have been previous studies of the density state of Si(111)-(7$\times$7) surfaces both at room\cite{si_sts1, si_sts2, si_sts3, PhysRevB.73.161302} and low\cite{PhysRevB.73.161302, Odobescu_2012} temperatures. 
Previous studies have reported that the Si surface has different electronic state behaviors depending on temperatures. 

The results suggest the importance of performing the site-specific measurement at room temperature. On the other hand, in the room temperature environment, in addition to the thermal drift and tip instability described above, there is also the effect of thermal fluctuations in the LDOS itself due to the broadening of the Fermi function for the population of electrons, resulting in a reduction in the energy resolution of the STS measurement. 
Therefore, in order to obtain a reliable site-specific measurement, it is necessary to acquire a large amount of data and statistically process the acquired data to deal with irregularities such as data variations and probe changes specific to room temperature.
For site-specific measurements at room temperature, the acquisition of large amounts of data has been very difficult with previous room temperature STM setups.

To ensure the acquisition of reliable data in this inherently uncertain environment, we've applied our AI-SPM to perform a large number of $I-V$ curve measurements on the four different adatom sites of the Si(111)-(7$\times$7) and performed statistical analysis to calculate the STS data.
Specifically, it can locate individual atomic sites of the respective center and corner adatoms of each faulted and unfaulted unit cell where adsorptions and defects were absent. 

\begin{figure}[h]
\centering
\includegraphics[width=0.8\linewidth]{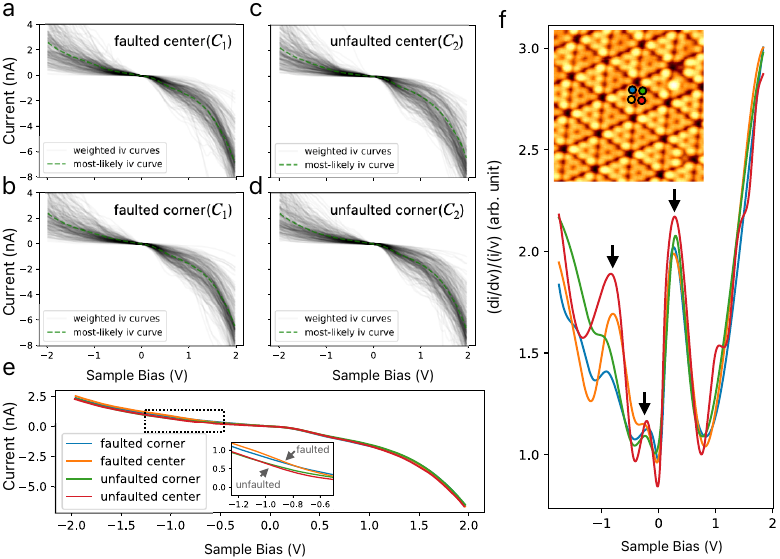}
\caption{
$I-V$ and calculated STS curves on different adatom site of the Si(111)-(7$\times$7) measuered at room temperature.
Weighted plot of $I-V$ curves on (a) faulted center adatom (b) faulted corner adatom (c) unfaulted center adatom (d) unfaulted corner adatom. The green dashed lines are the most likely ones calculated by the overall data-based selection method. 
(e) Over-wrapped $I-V$ curves of the four different adatom sites that are the same as the green dashed curves in (a) to (d).
(f) STS ($dI/dV$) curves calculated from the four $I-V$ curves in (e).
The inset in (f) is STM topographic image acquired with the -1.4~V sample bias, 200~pA tunneling current and the scan range is 11.25~nm$^2$. 
The four points on the topography represent the $I-V$ measurement position.
The color in the topography points and $dI/dV$ line correspond to each other. 
the measurement noise is reduced by the Savitzky-Golay and mean filters.
}
\label{fig:si_sts}
\end{figure}

Using this approach, we acquired a total of 324 $I-V$ curves measured at center and corner adatoms within both faulted and unfaulted unit cells, respectively. 
Following these measurements, the tip apex condition is evaluated to ensure that it remains optimal, and then the thermal drift is corrected again. 
$I-V$ data were automatically obtained in appropriate regions and tip conditions as evaluated by AI-SPM.
This entire procedure is performed iteratively, enabling the generation of robust and reliable data under the challenges of room temperature.

Obtained $I-V$ curves are individually plotted in Fig.~\ref{fig:si_sts}(a), (b), (c), and (d).
Our preliminary experiments (see Appendix B) have shown that the state of the tip apex changes about 7\% of the time when the voltage is swept.
To determine the most representative values for the $I-V$ curves of individual atoms (Fig.~\ref{fig:si_sts}(e)), we employed the data selection approach to obtain a group of curves with the highest tendencies, and the mean $I-V$ curves of the extracted group (see Methods).
$dI/dV$ curves were calculated by averaging the extracted $I-V$ curves followed by the mean filter, Savitzky-Golay\cite{savgol} filter, and smoothing by the 1D spline function to minimize the noise.
When zooming in on the region with negative sample bias, subtle differences between the curves of faulted unit cells and unfaulted unit cells become apparent. 
In this region, the conductance of the faulted unit cells is noted to be greater than that of the unfaulted unit cells, supporting the higher electron density on the faulted side with a stacking fault than on the unfaulted side\cite{PhysRevB.34.1388}.

As shown in the inset of Fig.~\ref{fig:si_sts}(f), in the STM measurements of the Si(111) surface, the contrast between the corner adatom and the center adatom in the faulted and unfaulted half-unit cells is different in the negative sample bias region. 
The calculated STS results also show that, under the negative sample bias region, the four site-specific locations at faulted/unfaulted corner and center adatom show distinct variations (the color of the points representing the positions of the four atoms corresponds to the color of the $dI/dV$ curves).
Because of the charge transfer from adatom to rest atom, the center adatom, which has a larger number of neighbors to the rest adatom, has a lower electron density than the corner adatom\cite{si_sts3}.
This trend of the four sites is revealed in our STS curves, and the positions of peaks [downward arrow in Fig.~\ref{fig:si_sts}(f)] align with the past result\cite{PhysRevB.73.161302}. 
Additionally, the most prominent peaks on the surface states of Si(111)-($7\times 7$) appear at -0.8~V and 0.3~V, supported by the ultraviolet photoemission spectroscopy and inverse photoemission spectroscopy measurement \cite{si_ups, si_ips}. 
Comparing the difference among site-specific STS curves, it is evident that the density state of the center adatom contributes the most to the -0.8~V peak on the surface states. 
By employing statistical analysis considering data variance in big data, which is obtained by an AI-assisted measurement approach to expand the IV curve to a data set, our method has shown a demonstration of the room temperature measurement with reliability and validity.

\section*{Discussion}

Our research shows that the AI-SPM system has the potential to revolutionize room temperature SPM by addressing longstanding challenges and opening new avenues for materials characterization. This implementation can also be used in variable temperature STM as well as cryogenic SPM. The thermal correction will be less relevant for the cryogenic environment but the automation part can be used. In fact, even in cryogenic environments, in experiments over the span of days, thermal drift will occur and the tip of the probe will change. For site-specific experiments over a long period of time, AI-SPM will be definitely a powerful tool in cryogenic environments.

In room temperature environments, various heat-assisted processes, including diffusion, crystal growth, dislocation movements, and chemical reactions, occur at the nano- to atomic-scale. 
Precise site-specific measurements could potentially unveil new scientific domains if local conditions are accurately assessed. 
However, thermal drift and tip apex change in SPM often obstruct the precise measurements. 
Furthermore, the necessity to collect and statistically analyze extensive datasets in heat-present environments presents a substantial challenge, often beyond the capacity of manual operations. 
Previously, researchers have achieved excellent results with diffusion and atomic manipulation using room temperature SPM, but it required patience, time, and luck.

To overcome these difficulties at room temperature, we have shown that a deep learning-assisted AI-SPM system capable of autonomously collecting real-time data at room temperature can be deployed.
A neural network is trained for comprehensive Si(111)-(7$\times$7) topography assessment, achieving an impressive accuracy at around  $90\%$, thanks to our two-phase data acquisition scheme which can automatically collect training datasets in real-time measurement. 
Moreover, this dataset acquisition routine can be applied to various surfaces, allowing users to train their own AI models for experimentation.

This transformation is evident not only in its capacity to assist humans in accomplishing complex and time-consuming operations, thus enabling automated experiments but also in the extension of SPM to the big data aiming to unlock deeper physical discoveries, as we delve into the analysis of a vast amount of data. In the future, it could be used effectively for SPM analysis at the atomic to nanometer level in materials with high temperatures and temperature variations, such as vanadium dioxide and thermoelectric materials. To achieve this vision, we need higher-bandwidth, automated SPM systems coupled with AI and real-time data analysis algorithms to provide intelligent data that is directly relevant to experimental conditions and scan results. 
However, there is a limitation of our current system, which demands preliminary experiments for executing the automatic data acquisition routine and gathering the training dataset necessary for the automation AI.
In future developments, training large models such as Vision Transformer\cite{dosovitskiy2021image}, capable of encompassing a wide range of material structures would enable the utilization of neural networks for the analysis of general SPM data.

This manuscript has introduced the concept of a self-driving lab, setting the stage for the development of innovative approaches, such as the automation and optimization of manufacturing processes, and the operation, self-replication, and self-repair of molecular machines in the realm of atomic technology. These advancements align closely with Drexler's anticipated integration of nanotechnology and AI\cite{Drexler1986Engines}. 

\section*{Conclusions}

We demonstrated the effectiveness of the AI-SPM system through two automated scanning experiments conducted at room temperature. We implemented an AI-equipped SPM capable of compensating for thermal drift and repairing tip damage, which have been challenges specific to the room-temperature operation. The system automatically addressed the challenge of room temperature by locating adsorbate-free regions and acquiring a large number of $I-V$ curves at four different adatom sites of the Si(111)-(7$\times$7) surface. Our results indicated a 6.3\% probability of changes occurring in the probe tip during the $I-V$ measurement process. This suggests that for reliable measurements at room temperature, it is necessary to acquire a significant amount of big data and subject these data to statistical processing. Such a process was not feasible with conventional SPM controllers for data acquisition, highlighting the need for AI support. STS measurements on the Si(111)-(7$\times$7)) demonstrated experimentally that the electronic state varies across four different adatom sites, confirming that STS characterization can be effectively performed at room temperature.

%\backmatter

\section*{Acknowledgements}

This work was supported in-part by a Grant-in-Aid for Scientific Research (19H05789, 21H01812, 22K18945) from the Ministry of Education, Culture, Sports, Science and Technology of Japan (MEXT).

\section*{Methods}

\subsection*{Sample preparation and experiment environment}

All experiments were carried out using a home-built STM operated at room temperature under ultrahigh vacuum conditions ($<\times10^{-8}$~Pa). 
The experiments were repeated across multiple sessions with different Pt/Ir STM tips to ensure reproducibility.
A n-type low-doped ($\rho \leq 0.02\ \Omega$cm) Si substrate is used in this research. 
Atomically flat and clean Si(111)-(7$\times$7) surfaces were prepared with the standard cleaning procedure, and used for dataset acquisition and experiment demonstration. 

For data acquisition, we have established SPM system augmented with a deep learning model explicitly designed for facilitating autonomous measurements, as depicted in Fig.~\ref{fig:aispm}. 
Our STM implementation combines an SPM instrument with a server that includes an ``AI Inference" subsystem for deep learning prediction, and a real-time operating subsystem with ``Scan Module" blocks that remotely control the SPM hardware [Fig.~\ref{fig:aispm}]. 
The control unit of the SPM instrument is built on an FPGA with remote access from a PC. 
The system was built with LabVIEW, LabVIEW FPGA, and Python. The scanning and data acquisition methods were performed in Python. NI PXIe-7857r was used as the measurement board.
The Scan Module block in the server contains SPM automation functions for optimizing the measurement environment. 
It encompassed the scan operation with custom external scripts in Python that automated SPM measurement routines and contained scan functions that optimized the experimental environment. 
OpenCV\cite{opencv_library} and SPMUtil\cite{spmu} python packages were employed for data processing and image processing. 
Communication between the AI Inference subsystem and the SPM instrument exploited a TCP protocol connection. 
AI Inference subsystem runs on Python and PyTorch\cite{pytorch} is adapted as the machine learning framework to perform training and inference, which is accelerated by RTX 4090 GPU for tensor computation.

\subsection*{Thermal drift compensation}

The real-time thermal drift compensation is based on an algorithm that extracts and matches feature points across consecutive scanned images\cite{drift}.   
By computing pixel shifts between consecutive scan images, the scan area is offset to track the original region corresponding to the first image. 
Furthermore, the drift velocity along the $x$, $y$, and $z$ is calculated from inter-image shifts and the data acquisition time and is utilized as the real-time drift compensation speed using a feedforward technique\cite{drift1, drift2}. 
The compensation process proceeded iteratively, acquiring images and adjusting for drift until the measured drift fell below a 0.2~nm threshold. 
Drift compensation is enabled to update the drift velocity at an interval of 10~min after the previous compensation is completed.

\subsection*{Tip apex optimization}

A protocol of tip apex optimization is implemented to modify and evaluate tip apex conditions through controlled mechanical impacts\cite{tipfix}. 
The bad tip is intentionally brought to poke toward the surface, inducing apex changes.
As a model case, the probe is indented 0.9~nm towards the surface relative to the 1.5~V sample bias and 200~pA tunneling current setpoint.
After poking, the subsequent scanned image is input to Net1 for tip quality determination. 
Unsatisfactory tip states based on the network output triggered additional pokes.
If a poke does not induce a change in the probe, the next poke will move the probe an additional 0.15 nm closer to the surface.
In this work, the $K_i (i=0,1,2,3)$ label in Fig. \ref{fig:model_architecture} (b) is regarded as a desired tip state for further experiments. 
The efficacy and responsibility of this automated optimization routine at room temperature have been demonstrated previously.

\subsection*{Training the deep learning model}

Neural network types and hyperparameters for training Net1, Net2, and Net3 used in this work are listed in Tab.~\ref{tab:train_param}.
Net1 uses a custom convolutional network structure which is presented in our previous study\cite{tipfix}.
Net2 and Net3 use the YOLOv8-small and YOLOv8-large structure\cite{yolov8} and the initial trainable parameters in the network are loaded from the pre-trained model. 
The YOLOv8-large in Net3 has more trainable parameters than YOLOv8-small and the larger model is chosen to improve the positioning accuracy on locating the atom key point.

%pre-processing
At the first stage for training Net1, Net2, and Net3, input images were pre-processed with the plane fit subtraction and Gaussian-Hann filters, and the size of the filtered images were adjusted to 256 x 256 pixels. The training dataset can be automatically labeled by the trained model. Subsequently, the labeled dataset is exported to CVAT for manual validation.
A balanced dataset was created, comprising more than 10,000 samples collected from experiments. 
Within this dataset, 2082 samples were designated for training Net1, 255 samples for training Net2, and 545 samples for training Net3.
These datasets were divided into 80\% for training and 20\% for validation.
Then, data augmentation\cite{albumentations}, which contained random affine transform, image coping, and contrast change, was separately applied to the training dataset and the validation dataset and expands the dataset by 5 times. 
%train
The three networks (Net1, Net2, and Net3) were trained by AdamW\cite{adamw} optimizer.

%\begin{sidewaystable}
%\begin{table}[ht]
%\centering
%\caption{\label{tab:train_param} The detail of the neural network models.}
%\begin{tabular}{*7c}
% \toprule 
%         \multicolumn{1}{c}{} 
%        &\multicolumn{3}{c}{model type } 
%        & \multicolumn{3}{c}{hyperparameter} \\
%        \cmidrule(lr){2-4}  \cmidrule(lr){5-7}
%Name & Architecture & Task & Output type & Learning rate & Training epoch & Batch size \\
% \midrule
%\hline
%Net1 & Custom CNN & tip \& sample classification & scalar  & $5\times 10^{-4}$ & 720 & 16 \\
%\hline
%Net2 & YOLOv8s & adsorption detection & bounding boxes & $2\times 10^{-3}$ & 150 & 16 \\
%\hline
%Net3 & YOLOv8l & atomic site detection & bounding boxes/key points & $1\times 10^{-3}$& 300 &16 \\
%\hline
%\end{tabular}
%\end{table}
%\end{sidewaystable}

\subsection*{$I-V$ curves similarity metric and selection}

To extract the high tendency from the part of the $I-V$ curves data group, the cosine similarity metric $\phi$ is used to process statics analysis.
The similarity $\phi$ of two $I-V$ curves $I_i, I_j\ (i,j = 1,2,\cdots, m)$ each containing $n$ count data points in one curve is calculated by the following equation. 

\begin{equation}
\label{eq:similarity}
	\phi (I_i, I_j) = \frac{\Sigma_{k=1}^n I_{i}[k] \cdot I_{j}[k]}{\sqrt{\Sigma_{k=1}^n  I_i[k]^2 \cdot \Sigma_{k=1}^n I_j[k]^2}}
\end{equation}

Especially, $\phi(I_1, I_2)$ is a range from -1 to 1, where $\phi(I_1, I_2) = 1$ means the two curve is fully equal.
By calculating the $\phi$ among all $I-V$ curves, we can obtain a group of curves that can represent the most common data of the whole data.
The major group selection method can be applied in two ways, which are based on the overall data (overall selection) or based on selected data (reference data-based selection), respectively. 
As for the overall selection method, given the whole $I-V$ curve data as $I_1, I_2,\cdots, I_l,\cdots, I_m$, we first calculate the mean value of cosine similarity ($\phi_{\mathrm{mean}}[I_l]$) for each $I-V$ curve responding to the other all curves.

\begin{equation}
\phi_{\mathrm{mean}}([I_l]) = \Sigma^i_m \phi([I_l, I_i]) / m
\end{equation}

The $\phi_{\mathrm{mean}}([I_l])$ for $I_l$ means the common property to the whole $I-V$ curves, so we then apply a threshold $T_1$ with the condition $ \phi_{\mathrm{mean}}([I_l]) > T_1$ to choose all $I_l$ curve that meets this condition as a major group. 
The reference data-based selection method is to select a group of data similar to the reference data, which is one of an $I-V$ curve from the $I-V$ curves group as the reference data $I_{\mathrm{ref}}$.
For an $I-V$ curve $I_l$, we give a threshold $T_2$ and use the $\phi(I_{\mathrm{ref}}, I_l) > T_2$ condition to judge if $I_l$ can be added into the major group which is based on $I_{\mathrm{ref}}$.
Before comparing the relationship between the $I-V$ curves, the measurement noise is reduced by the Savitzky-Golay\cite{savgol} and mean filters.
After the major group is extracted, The most proper tendency in the group can be extracted by taking the average value in the $I-V$ curves and smoothing by a 1D spline function.

\section{Model evaluation}\label{sup:model}

\begin{figure}[h]
\centering
\includegraphics[width=0.8\linewidth]{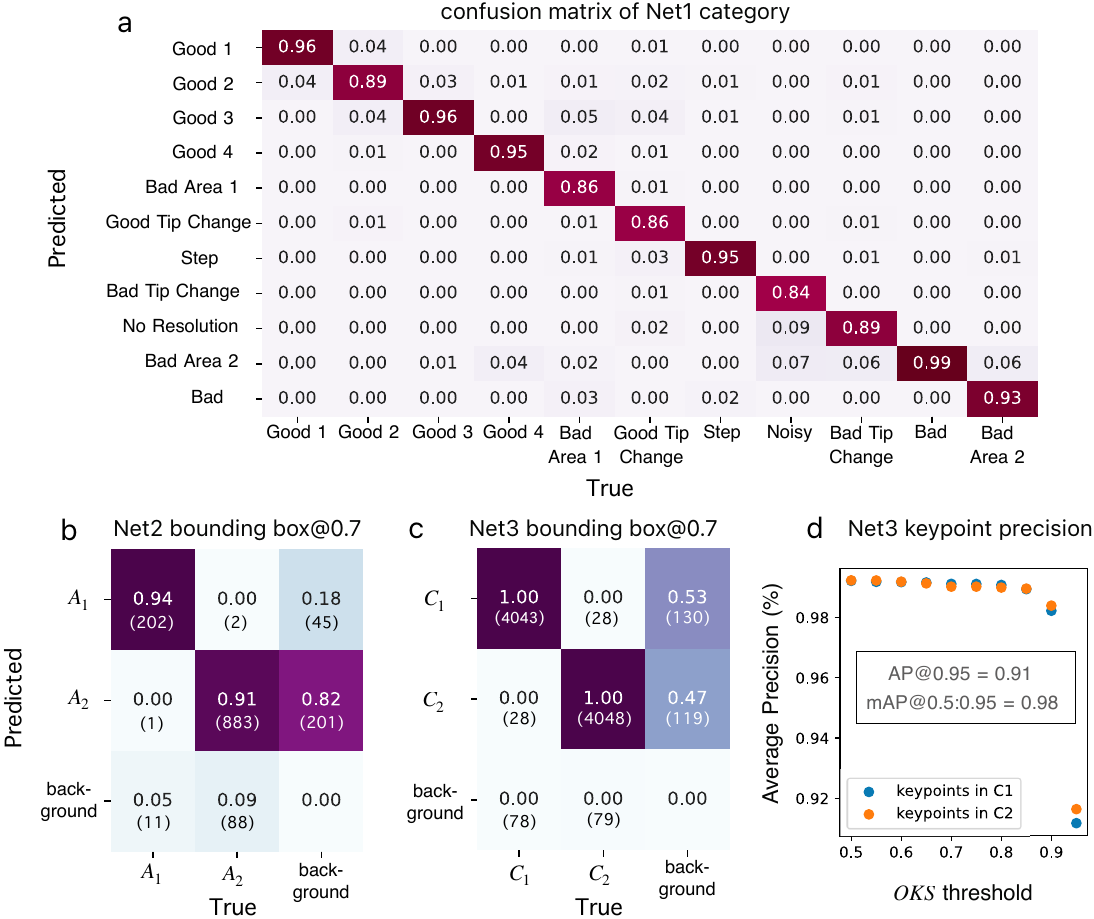}
\caption{
Evaluation of the CNN performance of the room temperature STM on the $\mathrm{Si(111)-(7\times 7)}$. 
(a) to (c) Confusion matrix demonstrating classification accuracy of Net1 to Net3, respectively.
In (b) and (c), background means AI does not treat the detection object as an interest to accurately identify, and value of $IoU\geq 0.7$ is defined as the true positive case. 
Numbers enclosed in brackets within the matrix elements, representing the number of test samples.
(d) $OKS$ threshold dependence of AP(Average Precision) analysis of Net3 for atom point detection. mAP calculates the mean value of the values of AP corresponding to the $OKS$ from 0.5 to 0.95. 
}
\label{fig:model_performance}
\end{figure}

Several metrics are introduced to assess model performance, specifically to validate the accuracy of classification and detection. For Net1, a confusion matrix is employed to verify classification accuracy as in Fig. \ref{fig:model_performance}(a).
In the matrix, each element represents the number of instances by authors (horizontal axis) and the Net1 predictions (vertical axis), and it is normalized by the number of ground truth instances for each class. The diagonal elements represent the {\it Recall} metric, defined as:

\begin{equation}
Recall = \frac{TP}{TP + FN},
\end{equation}

where true positive (TP) denotes the number of predictions identified correctly as positive among the actual positives, and false negative (FN) refers to the number of actual positives incorrectly predicted as negative. The Recall metric serves as an indicator of reproducibility when the model is applied within a measurement system. For instance, as illustrated by the first row and column in Fig. \ref{fig:model_performance}(a), scanning with a tip in the "Good 1" state provided the system with 0.98 likelihood of successful identification. The average {\it Recall} across all classes was 0.93, and the {\it Recall} for binary judgment, determining whether a tip was good or bad, was 0.98.
Compared to our previous study\cite{tipfix} in which the same algorithm of Net1 has been used, we have confirmed the improvement of the $Recall$ value of Net1 from 87\% to 93\%.
This enhancement is attributed to the expansion of the dataset to 2000.

For Net2, the performance of the moving adsorption ($A_1$), and non-moving adsorption ($A_2$) detection is validated. 
To define the success or failure of the detection, the Intersection Over Union($IoU$) metric which quantifies the overlap between the predicted bounding box and the ground truth bounding box for a given object instance, was introduced.
It is defined as the intersection area between the predicted box and ground truth box divided by the union of the two boxes,
\begin{equation}
	IoU = \frac{S_{\mathrm{overlap}}}{S_{\mathrm{union}}},
\end{equation}
where $S_{\mathrm{overlap}}$ is the area of overlap of the predicted and ground truth boxes, and $S_{\mathrm{union}}$ is their union area, which covered by both boxes.
We defined the threshold value of $IoU  = 0.7$ to determine if a prediction should have been considered a true or false positive. 
Specifically, a predicted box with $IoU \geq$ 0.7 concerning the ground truth is counted as a true positive (TP), while a false negative (FN) indicates a failure to detect a ground truth object. 
Using this categorization, a confusion matrix can then be constructed to summarize the performance of each object class on the validation datasets.
Fig. \ref{fig:model_performance} (b) shows the normalized confusion matrix when $IoU \geq 0.7$ was used as the threshold to divide the predictions in validation datasets.
The category label also contains “background”, which represents the object that does not contain any target. The reason that a certain amount of background is misidentified as a target object may be because our detection model will repeatedly detect specific targets. The average Recall of detecting $A_1$ and $A_2$ is 0.92.

For Net3, Object Keypoint Similarity($OKS$) was introduced to verify the detection accuracy of the bounding box ($C_1$, $C_2$) and key points ($P_i$) as, 
\begin{equation}
OKS = \frac{\Sigma_i^n \mathrm{exp}(-d_i^2/2s^2)\delta(v_i>0)}{\Sigma_i^n\delta(v_i>0)},
\end{equation}
where $n$ is the number of key points, $d_i$ is the Euclidean distance between the predicted keypoint and its corresponding ground truth, $s$ is the region scale of the detected object, and $\delta(v_i>0)$ is an indicator function that resolves to 1 if keypoint is visible or 0 if occluded.
$OKS$ ranges from 0 to 1, with 1 indicating perfect alignment of predicted key points to the ground truth.
A threshold value of $OKS = 0.5$ was used to determine true versus false predicted key points and $OKS \geq$ 0.95 can give a rigorous evaluation precision of the atom localization quality.
The overall keypoint detection performance can then be quantified by the proportion of key points with $OKS$ above a threshold. 
This is referred to as the mean Average Precision (mAP) for one of the keypoint classes.

In Fig. \ref{fig:model_performance} (c), the normalized confusion matrix of two unit cells (C1, C2) and background is shown.
Here, similarly to Net2, $IoU = 0.7$ was chosen as the threshold to plot the normalized confusion matrix of two unit cells (C1, C2) and background.
A substantial quantity of unit cell training samples allowed the {\it Recall} value to approach its maximum performance level of 1.0.

Besides, the capability of accurately localizing atom keypoints can be evaluated by {\it OKS} value. When choosing {\it OKS} thresholds varying by 0.05 from 0.5 to 0.95, the respect of Average Precision detected in C1 and C2 are plotted in Fig. \ref{fig:model_performance} (d). mAP was above 0.98, and as the most rigorous metric, mAP value at $OKS \geq 0.95$ was 0.91.
Overall, these models are trained on the variety of the dataset which are acquired in real-time by a step-up data acquisition scheme. All of the models achieved high accuracy performance even for real-time data at room temperature, thus ensuring the robustness of the measurements and localization performance for SPM automation.

\section{Statistical methods in room temperature $I-V$ measurements}\label{sup:iv}

\begin{figure}[h]
\centering
\includegraphics[width=0.95\linewidth]{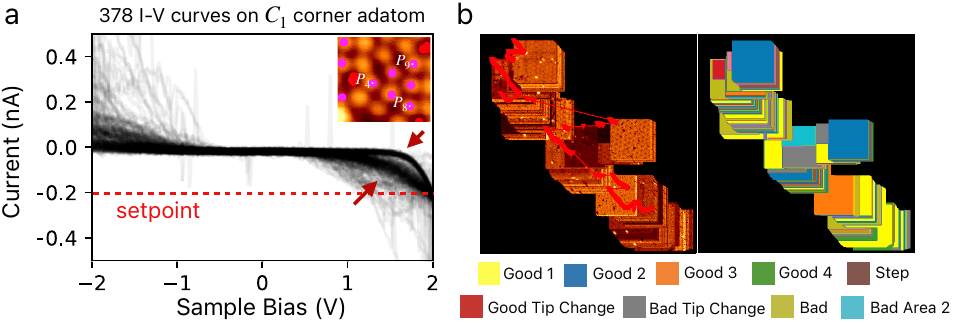}
\caption{
Site-specific $I-V$ measurement by AI-SPM on the $\mathrm{Si(111)-(7\times 7)}$ surface at room temperature. The STM tip was repaired while searching the region, and data were automatically acquired in the region deemed appropriate for the $I-V$ measurement.
(a) $I-V$ curves measured at the corner adatoms of faulted half-unit cells ($C_1$) on the $\mathrm{Si(111)-(7\times 7)}$ surface. A total of 378 curves are superimposed. The horizontal red dashed line indicates the set point of the tunneling current before the start of the IV scan.  "$P_4$", "$P_8$", and "$P_9$" in the inset are corner adatoms within $C_1$ identified by Net3, where the $I-V$ curves were automatically measured. In reality, measurements were acquired at four different atomic sites on Si(111).
(b) Top-most inferred state of the STM images during the self-driving $I-V$ measurements by our AI-SPM. The notation is the same as in Fig. \ref{fig:demo1_auto_scan}(b).
}
\label{fig:demo2_sts_scan}
\end{figure}

\begin{figure}[h]
\centering
\includegraphics[width=0.8\linewidth]{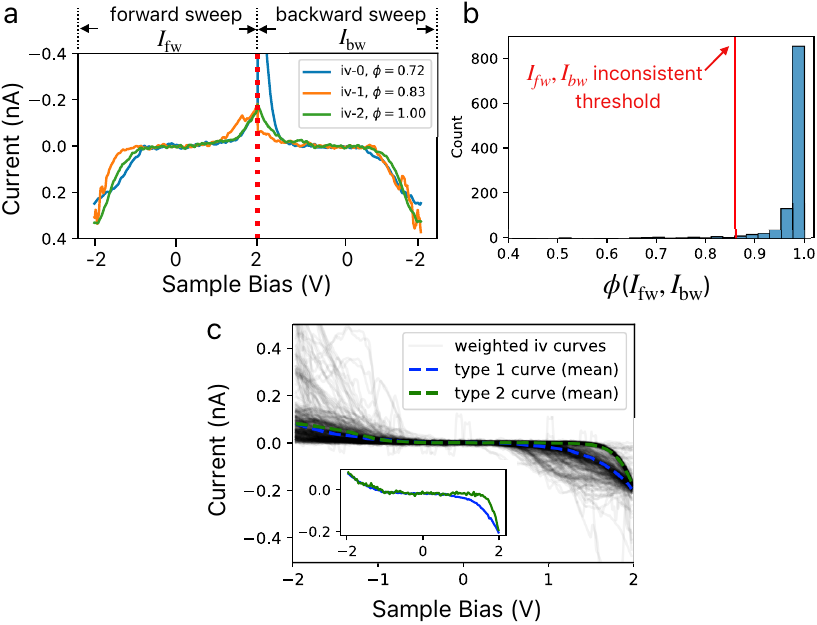}
\caption{
Statistical analysis on 378 curves on $C_1$ corner adatom. 
(a) Three raw data("iv-0", "iv-1", "iv-2") of the IV measurement, which is taken from setting sample bias from $2$~V to $-2$~V as the forward sweep and then $-2$~V to $2$~V as the backward sweep, starting from $-200$~pA setpoint at $2$~V sample bias. 
(b) Histogram of the similarity metric $\phi(I_{\mathrm{fw}}, I_\mathrm{{bw}})$ between forward $I-V$ curve $I_\mathrm{{fw}}$ and backward $I-V$ curve $I_\mathrm{{bw}}$.
(c) The result of two $I-V$ curves which represent the two tend group's mean value. Type 1 curve is selected by the overall data-based selection method and type 2 curve is selected by the reference curve-based selection method (see Methods).
}
\label{fig:analysis}
\end{figure}

Figure \ref{fig:demo2_sts_scan} shows the results of a long-time site-specific $I-V$ measurement, performed to verify the rate of change of the probe. The measurement is the same as in Fig. \ref{fig:si_sts}, but a different STM tip was used.
The measurement took 58 hours and totally 695 topographic STM images and 2832 $I-V$ curves were obtained.
The selected 378 $I-V$ curves acquired on corner atoms of the half unit cell (the 4th, 8th, and 9th points detected by Net3)  are shown in \ref{fig:demo2_sts_scan}(a). 
Trajectory of the imaging area and the top-most inferred state of the STM images during the self-driving $I-V$ measurements by our AI-SPM is shown Fig. \ref{fig:demo2_sts_scan}(b).
The $I-V$ curve measurement contains both a forward sweep ($I_{\mathrm{fw}}$), ranging from $2$~V to $-2$~V of sample bias, and a backward sweep ($I_\mathrm{{bw}}$) from $-2$~V to $2$~V of sample bias. 
The scan is initiated from the target atom where the $z$ position is specifically at a $-200$~pA setpoint on a $2$~V sample bias. 
IV curves in Fig.\ref{fig:analysis}(a) encompass both $I_{\mathrm{fw}}$ and $I_{\mathrm{bw}}$ and almost all curves precisely align at the $-200$~pA setpoint under a $2$~V sample bias position. 
This observation indicates that drift along the $x$, $y$, and $z$ axes has been effectively corrected, offering compelling evidence that the system can mitigate the impact of thermal drift even during room temperature measurements.

However, due to various environment changes during the measurement process, the numerical values of $I_\mathrm{{fw}}$ and $I_\mathrm{{bw}}$ may not always align.
In cases where the tip stabilizes, and the surface remains undisturbed, $I_\mathrm{{fw}}$ and $I_\mathrm{{bw}}$ tend to be consistent. 
To assess the level of agreement between $I_\mathrm{{fw}}$ and $I_\mathrm{{bw}}$ curves, we used the cosine similarity metric (see Method).
In Fig.~\ref{fig:analysis}(a), as examples, three curves (iv-0, iv-1, and iv-2) are shown. As in the case of iv-0 and iv-1, respective $\phi$ values of 0.72 and 0.83 are evident that tip apex change happened.
In contrast, when there is a high degree of consistency, as observed in the curve of "iv-2," the curve achieves a $\phi$ value of 1.00.

All $\phi(I_\mathrm{{fw}}, I_\mathrm{{bw}})$ values across a set of $I-V$ curves are computed and depicted in the histogram in Fig.~\ref{fig:analysis}(b).
We defined $\phi(I_\mathrm{{fw}}, I_\mathrm{{bw}}) < 0.86$ as indicating a flaw in $I-V$ measurements caused by changes in measurement conditions, the probability of encountering such measurement discrepancies using our STS parameter was $6.3\%$. 
The reason behind these discrepancies is often associated with the impact on the tip apex resulting from the applied sample bias, which can be likened to a process of conditioning the probe\cite{ai_sts}. 
These cases are frequently observed during measurements, manifesting as tip changes or impurities dropping onto the surface, and are largely unavoidable when operating at room temperature.

Despite the potential influence of an unstable measurement environment as described earlier, trends in surface properties can still be observed, especially when supported by adequate data. 
Considering that all the curves have been weighted in the plot, the areas with denser line colors indicate higher-weight regions, reflecting a more pronounced tendency in the $I-V$ curve within those specific regions.
Within the range of 1 to 2~V Sample Bias, two $I-V$ curves exhibit the most prominent directions.
To distinguish and represent these two distinct trends, an algorithm based on cosine similarity (see Method) is employed to statistically differentiate and select the sets of $I-V$ curves representing these two tendencies. 
The average values of all curves within each set, corresponding to the two tendencies, are plotted as blue and green dotted lines in Fig.~\ref{fig:analysis}(c). 
The reason behind the differentiation of these two trends on corner adatoms of the $C_1$ type, despite their identical nature, might be attributed to variations in the density of the state of the STM tip apex during $I-V$ curve measurements.
Though our automated SPM system guarantees the ability to acquire atomic-resolution topography, it still needs to be improved to maintain consistent local density of states of the tip during measurements.

\bibliography{ref}

%aipnum4-2.bst 2019-01-14 (MD) hand-edited version of apsrev4-1.bst
\begin{thebibliography}{61}%
\makeatletter
\providecommand \@ifxundefined [1]{%
 \@ifx{#1\undefined}
}%
\providecommand \@ifnum [1]{%
 \ifnum #1\expandafter \@firstoftwo
 \else \expandafter \@secondoftwo
 \fi
}%
\providecommand \@ifx [1]{%
 \ifx #1\expandafter \@firstoftwo
 \else \expandafter \@secondoftwo
 \fi
}%
\providecommand \natexlab [1]{#1}%
\providecommand \enquote  [1]{``#1''}%
\providecommand \bibnamefont  [1]{#1}%
\providecommand \bibfnamefont [1]{#1}%
\providecommand \citenamefont [1]{#1}%
\providecommand \href@noop [0]{\@secondoftwo}%
\providecommand \href [0]{\begingroup \@sanitize@url \@href}%
\providecommand \@href[1]{\@@startlink{#1}\@@href}%
\providecommand \@@href[1]{\endgroup#1\@@endlink}%
\providecommand \@sanitize@url [0]{\catcode `\\12\catcode `\$12\catcode
  `\&12\catcode `\#12\catcode `\^12\catcode `\_12\catcode `\%12\relax}%
\providecommand \@@startlink[1]{}%
\providecommand \@@endlink[0]{}%
\providecommand \url  [0]{\begingroup\@sanitize@url \@url }%
\providecommand \@url [1]{\endgroup\@href {#1}{\urlprefix }}%
\providecommand \urlprefix  [0]{URL }%
\providecommand \Eprint [0]{\href }%
\providecommand \doibase [0]{https://doi.org/}%
\providecommand \selectlanguage [0]{\@gobble}%
\providecommand \bibinfo  [0]{\@secondoftwo}%
\providecommand \bibfield  [0]{\@secondoftwo}%
\providecommand \translation [1]{[#1]}%
\providecommand \BibitemOpen [0]{}%
\providecommand \bibitemStop [0]{}%
\providecommand \bibitemNoStop [0]{.\EOS\space}%
\providecommand \EOS [0]{\spacefactor3000\relax}%
\providecommand \BibitemShut  [1]{\csname bibitem#1\endcsname}%
\let\auto@bib@innerbib\@empty
%</preamble>
\bibitem [{\citenamefont {Drexler}(1986)}]{Drexler1986Engines}%
  \BibitemOpen
  \bibfield  {author} {\bibinfo {author} {\bibfnamefont {K.~E.}\ \bibnamefont
  {Drexler}},\ }\href@noop {} {\emph {\bibinfo {title} {Engines of Creation:
  The Coming Era of Nanotechnology}}}\ (\bibinfo  {publisher} {Anchor
  Press/Doubleday},\ \bibinfo {address} {Garden City, NY},\ \bibinfo {year}
  {1986})\BibitemShut {NoStop}%
\bibitem [{\citenamefont {Binnig}\ \emph {et~al.}(1982)\citenamefont {Binnig},
  \citenamefont {Rohrer}, \citenamefont {Gerber},\ and\ \citenamefont
  {Weibel}}]{stm}%
  \BibitemOpen
  \bibfield  {author} {\bibinfo {author} {\bibfnamefont {G.}~\bibnamefont
  {Binnig}}, \bibinfo {author} {\bibfnamefont {H.}~\bibnamefont {Rohrer}},
  \bibinfo {author} {\bibfnamefont {C.}~\bibnamefont {Gerber}},\ and\ \bibinfo
  {author} {\bibfnamefont {E.}~\bibnamefont {Weibel}},\ }\href
  {https://doi.org/10.1103/PhysRevLett.49.57} {\bibfield  {journal} {\bibinfo
  {journal} {Phys. Rev. Lett.}\ }\textbf {\bibinfo {volume} {49}},\ \bibinfo
  {pages} {57--61} (\bibinfo {year} {1982})}\BibitemShut {NoStop}%
\bibitem [{\citenamefont {Binnig}, \citenamefont {Quate},\ and\ \citenamefont
  {Gerber}(1986)}]{afm}%
  \BibitemOpen
  \bibfield  {author} {\bibinfo {author} {\bibfnamefont {G.}~\bibnamefont
  {Binnig}}, \bibinfo {author} {\bibfnamefont {C.~F.}\ \bibnamefont {Quate}},\
  and\ \bibinfo {author} {\bibfnamefont {C.}~\bibnamefont {Gerber}},\ }\href
  {https://doi.org/10.1103/PhysRevLett.56.930} {\bibfield  {journal} {\bibinfo
  {journal} {Phys. Rev. Lett.}\ }\textbf {\bibinfo {volume} {56}},\ \bibinfo
  {pages} {930--933} (\bibinfo {year} {1986})}\BibitemShut {NoStop}%
\bibitem [{\citenamefont {Eigler}\ and\ \citenamefont
  {Schweizer}(1990)}]{Eigler1990}%
  \BibitemOpen
  \bibfield  {author} {\bibinfo {author} {\bibfnamefont {D.~M.}\ \bibnamefont
  {Eigler}}\ and\ \bibinfo {author} {\bibfnamefont {E.~K.}\ \bibnamefont
  {Schweizer}},\ }\bibfield  {title} {\enquote {\bibinfo {title} {Positioning
  single atoms with a scanning tunnelling microscope},}\ }\href
  {https://doi.org/10.1038/344524a0} {\bibfield  {journal} {\bibinfo  {journal}
  {Nature}\ }\textbf {\bibinfo {volume} {344}},\ \bibinfo {pages} {524--526}
  (\bibinfo {year} {1990})}\BibitemShut {NoStop}%
\bibitem [{\citenamefont {Jel{\'\i}nek}(2017)}]{chem1}%
  \BibitemOpen
  \bibfield  {author} {\bibinfo {author} {\bibfnamefont {P.}~\bibnamefont
  {Jel{\'\i}nek}},\ }\href {https://doi.org/10.1088/1361-648X/aa76c7}
  {\bibfield  {journal} {\bibinfo  {journal} {Journal of Physics: Condensed
  Matter}\ }\textbf {\bibinfo {volume} {29}},\ \bibinfo {pages} {343002}
  (\bibinfo {year} {2017})}\BibitemShut {NoStop}%
\bibitem [{\citenamefont {Barth}\ and\ \citenamefont
  {Reichling}(2001)}]{chem2}%
  \BibitemOpen
  \bibfield  {author} {\bibinfo {author} {\bibfnamefont {C.}~\bibnamefont
  {Barth}}\ and\ \bibinfo {author} {\bibfnamefont {M.}~\bibnamefont
  {Reichling}},\ }\href@noop {} {\bibfield  {journal} {\bibinfo  {journal}
  {Nature}\ }\textbf {\bibinfo {volume} {414}},\ \bibinfo {pages} {54--57}
  (\bibinfo {year} {2001})}\BibitemShut {NoStop}%
\bibitem [{\citenamefont {Kodera}\ \emph {et~al.}(2010)\citenamefont {Kodera},
  \citenamefont {Yamamoto}, \citenamefont {Ishikawa},\ and\ \citenamefont
  {Ando}}]{hsafm}%
  \BibitemOpen
  \bibfield  {author} {\bibinfo {author} {\bibfnamefont {N.}~\bibnamefont
  {Kodera}}, \bibinfo {author} {\bibfnamefont {D.}~\bibnamefont {Yamamoto}},
  \bibinfo {author} {\bibfnamefont {R.}~\bibnamefont {Ishikawa}},\ and\
  \bibinfo {author} {\bibfnamefont {T.}~\bibnamefont {Ando}},\ }\href
  {https://doi.org/10.1038/nature09450} {\bibfield  {journal} {\bibinfo
  {journal} {Nature}\ }\textbf {\bibinfo {volume} {468}},\ \bibinfo {pages}
  {72--76} (\bibinfo {year} {2010})}\BibitemShut {NoStop}%
\bibitem [{\citenamefont {Jelic}\ \emph {et~al.}(2017)\citenamefont {Jelic},
  \citenamefont {Iwaszczuk}, \citenamefont {Nguyen}, \citenamefont {Rathje},
  \citenamefont {Hornig}, \citenamefont {Sharum}, \citenamefont {Hoffman},
  \citenamefont {Freeman},\ and\ \citenamefont {Hegmann}}]{dynamics1}%
  \BibitemOpen
  \bibfield  {author} {\bibinfo {author} {\bibfnamefont {V.}~\bibnamefont
  {Jelic}}, \bibinfo {author} {\bibfnamefont {K.}~\bibnamefont {Iwaszczuk}},
  \bibinfo {author} {\bibfnamefont {P.~H.}\ \bibnamefont {Nguyen}}, \bibinfo
  {author} {\bibfnamefont {C.}~\bibnamefont {Rathje}}, \bibinfo {author}
  {\bibfnamefont {G.~J.}\ \bibnamefont {Hornig}}, \bibinfo {author}
  {\bibfnamefont {H.~M.}\ \bibnamefont {Sharum}}, \bibinfo {author}
  {\bibfnamefont {J.~R.}\ \bibnamefont {Hoffman}}, \bibinfo {author}
  {\bibfnamefont {M.~R.}\ \bibnamefont {Freeman}},\ and\ \bibinfo {author}
  {\bibfnamefont {F.~A.}\ \bibnamefont {Hegmann}},\ }\bibfield  {title}
  {\enquote {\bibinfo {title} {Ultrafast terahertz control of extreme tunnel
  currents through single atoms on a silicon surface},}\ }\href
  {https://doi.org/10.1038/nphys4047} {\bibfield  {journal} {\bibinfo
  {journal} {Nature Physics}\ }\textbf {\bibinfo {volume} {13}},\ \bibinfo
  {pages} {591--598} (\bibinfo {year} {2017})}\BibitemShut {NoStop}%
\bibitem [{\citenamefont {Yoshida}\ \emph {et~al.}(2012)\citenamefont
  {Yoshida}, \citenamefont {Terada}, \citenamefont {Oshima}, \citenamefont
  {Takeuchi},\ and\ \citenamefont {Shigekawa}}]{dynamics2}%
  \BibitemOpen
  \bibfield  {author} {\bibinfo {author} {\bibfnamefont {S.}~\bibnamefont
  {Yoshida}}, \bibinfo {author} {\bibfnamefont {Y.}~\bibnamefont {Terada}},
  \bibinfo {author} {\bibfnamefont {R.}~\bibnamefont {Oshima}}, \bibinfo
  {author} {\bibfnamefont {O.}~\bibnamefont {Takeuchi}},\ and\ \bibinfo
  {author} {\bibfnamefont {H.}~\bibnamefont {Shigekawa}},\ }\bibfield  {title}
  {\enquote {\bibinfo {title} {Nanoscale probing of transient carrier dynamics
  modulated in a gaas--pin junction by laser-combined scanning tunneling
  microscopy},}\ }\href@noop {} {\bibfield  {journal} {\bibinfo  {journal}
  {Nanoscale}\ }\textbf {\bibinfo {volume} {4}},\ \bibinfo {pages} {757--761}
  (\bibinfo {year} {2012})}\BibitemShut {NoStop}%
\bibitem [{\citenamefont {Brihuega}, \citenamefont {Custance},\ and\
  \citenamefont {G\'omez-Rodr\'{\i}guez}(2004)}]{BrihuegaGeDiffusion}%
  \BibitemOpen
  \bibfield  {author} {\bibinfo {author} {\bibfnamefont {I.}~\bibnamefont
  {Brihuega}}, \bibinfo {author} {\bibfnamefont {O.}~\bibnamefont {Custance}},\
  and\ \bibinfo {author} {\bibfnamefont {J.~M.}\ \bibnamefont
  {G\'omez-Rodr\'{\i}guez}},\ }\href
  {https://doi.org/10.1103/PhysRevB.70.165410} {\bibfield  {journal} {\bibinfo
  {journal} {Phys. Rev. B}\ }\textbf {\bibinfo {volume} {70}},\ \bibinfo
  {pages} {165410} (\bibinfo {year} {2004})}\BibitemShut {NoStop}%
\bibitem [{\citenamefont {Custance}\ \emph
  {et~al.}(2001{\natexlab{a}})\citenamefont {Custance}, \citenamefont
  {Brihuega}, \citenamefont {G\'{o}mez-Rodr\'{i}guez},\ and\ \citenamefont
  {Bar\'{o}}}]{SnOnSi}%
  \BibitemOpen
  \bibfield  {author} {\bibinfo {author} {\bibfnamefont {O.}~\bibnamefont
  {Custance}}, \bibinfo {author} {\bibfnamefont {I.}~\bibnamefont {Brihuega}},
  \bibinfo {author} {\bibfnamefont {J.}~\bibnamefont
  {G\'{o}mez-Rodr\'{i}guez}},\ and\ \bibinfo {author} {\bibfnamefont
  {A.}~\bibnamefont {Bar\'{o}}},\ }\href
  {https://doi.org/https://doi.org/10.1016/S0039-6028(01)00732-4} {\bibfield
  {journal} {\bibinfo  {journal} {Surface Science}\ }\textbf {\bibinfo {volume}
  {482-485}},\ \bibinfo {pages} {1406--1412} (\bibinfo {year}
  {2001}{\natexlab{a}})}\BibitemShut {NoStop}%
\bibitem [{\citenamefont {Meusel}\ \emph {et~al.}(2021)\citenamefont {Meusel},
  \citenamefont {Gezmis}, \citenamefont {Jaekel}, \citenamefont {Lexow},
  \citenamefont {Bayer}, \citenamefont {Maier},\ and\ \citenamefont
  {Steinr{\"u}ck}}]{surface_growth}%
  \BibitemOpen
  \bibfield  {author} {\bibinfo {author} {\bibfnamefont {M.}~\bibnamefont
  {Meusel}}, \bibinfo {author} {\bibfnamefont {A.}~\bibnamefont {Gezmis}},
  \bibinfo {author} {\bibfnamefont {S.}~\bibnamefont {Jaekel}}, \bibinfo
  {author} {\bibfnamefont {M.}~\bibnamefont {Lexow}}, \bibinfo {author}
  {\bibfnamefont {A.}~\bibnamefont {Bayer}}, \bibinfo {author} {\bibfnamefont
  {F.}~\bibnamefont {Maier}},\ and\ \bibinfo {author} {\bibfnamefont {H.-P.}\
  \bibnamefont {Steinr{\"u}ck}},\ }\href@noop {} {\bibfield  {journal}
  {\bibinfo  {journal} {The Journal of Physical Chemistry C}\ }\textbf
  {\bibinfo {volume} {125}},\ \bibinfo {pages} {20439--20449} (\bibinfo {year}
  {2021})}\BibitemShut {NoStop}%
\bibitem [{\citenamefont {Sugimoto}\ \emph {et~al.}(2005)\citenamefont
  {Sugimoto}, \citenamefont {Abe}, \citenamefont {Hirayama}, \citenamefont
  {Oyabu}, \citenamefont {Custance},\ and\ \citenamefont {Morita}}]{am1}%
  \BibitemOpen
  \bibfield  {author} {\bibinfo {author} {\bibfnamefont {Y.}~\bibnamefont
  {Sugimoto}}, \bibinfo {author} {\bibfnamefont {M.}~\bibnamefont {Abe}},
  \bibinfo {author} {\bibfnamefont {S.}~\bibnamefont {Hirayama}}, \bibinfo
  {author} {\bibfnamefont {N.}~\bibnamefont {Oyabu}}, \bibinfo {author}
  {\bibfnamefont {{\'O}.}~\bibnamefont {Custance}},\ and\ \bibinfo {author}
  {\bibfnamefont {S.}~\bibnamefont {Morita}},\ }\href
  {https://doi.org/10.1038/nmat1297} {\bibfield  {journal} {\bibinfo  {journal}
  {Nature Materials}\ }\textbf {\bibinfo {volume} {4}},\ \bibinfo {pages}
  {156--159} (\bibinfo {year} {2005})}\BibitemShut {NoStop}%
\bibitem [{\citenamefont {Sugimoto}\ \emph
  {et~al.}(2008{\natexlab{a}})\citenamefont {Sugimoto}, \citenamefont {Pou},
  \citenamefont {Custance}, \citenamefont {Jel\'{i}nek}, \citenamefont {Abe},
  \citenamefont {P\'{e}rez},\ and\ \citenamefont {Morita}}]{am2}%
  \BibitemOpen
  \bibfield  {author} {\bibinfo {author} {\bibfnamefont {Y.}~\bibnamefont
  {Sugimoto}}, \bibinfo {author} {\bibfnamefont {P.}~\bibnamefont {Pou}},
  \bibinfo {author} {\bibfnamefont {O.}~\bibnamefont {Custance}}, \bibinfo
  {author} {\bibfnamefont {P.}~\bibnamefont {Jel\'{i}nek}}, \bibinfo {author}
  {\bibfnamefont {M.}~\bibnamefont {Abe}}, \bibinfo {author} {\bibfnamefont
  {R.}~\bibnamefont {P\'{e}rez}},\ and\ \bibinfo {author} {\bibfnamefont
  {S.}~\bibnamefont {Morita}},\ }\href@noop {} {\bibfield  {journal} {\bibinfo
  {journal} {Science}\ }\textbf {\bibinfo {volume} {322}},\ \bibinfo {pages}
  {413--417} (\bibinfo {year} {2008}{\natexlab{a}})},\ \Eprint
  {https://arxiv.org/abs/https://www.science.org/doi/pdf/10.1126/science.1160601}
  {https://www.science.org/doi/pdf/10.1126/science.1160601} \BibitemShut
  {NoStop}%
\bibitem [{\citenamefont {Abe}\ \emph {et~al.}(2005)\citenamefont {Abe},
  \citenamefont {Sugimoto}, \citenamefont {Custance},\ and\ \citenamefont
  {Morita}}]{drift1}%
  \BibitemOpen
  \bibfield  {author} {\bibinfo {author} {\bibfnamefont {M.}~\bibnamefont
  {Abe}}, \bibinfo {author} {\bibfnamefont {Y.}~\bibnamefont {Sugimoto}},
  \bibinfo {author} {\bibfnamefont {O.}~\bibnamefont {Custance}},\ and\
  \bibinfo {author} {\bibfnamefont {S.}~\bibnamefont {Morita}},\ }\bibfield
  {title} {\enquote {\bibinfo {title} {{Room-temperature reproducible spatial
  force spectroscopy using atom-tracking technique}},}\ }\href
  {https://doi.org/10.1063/1.2108112} {\bibfield  {journal} {\bibinfo
  {journal} {Applied Physics Letters}\ }\textbf {\bibinfo {volume} {87}},\
  \bibinfo {pages} {173503} (\bibinfo {year} {2005})}\BibitemShut {NoStop}%
\bibitem [{\citenamefont {Rahe}\ \emph {et~al.}(2011)\citenamefont {Rahe},
  \citenamefont {Sch{\"u}tte}, \citenamefont {Schniederberend}, \citenamefont
  {Reichling}, \citenamefont {Abe}, \citenamefont {Sugimoto},\ and\
  \citenamefont {K{\"u}hnle}}]{drift2}%
  \BibitemOpen
  \bibfield  {author} {\bibinfo {author} {\bibfnamefont {P.}~\bibnamefont
  {Rahe}}, \bibinfo {author} {\bibfnamefont {J.}~\bibnamefont {Sch{\"u}tte}},
  \bibinfo {author} {\bibfnamefont {W.}~\bibnamefont {Schniederberend}},
  \bibinfo {author} {\bibfnamefont {M.}~\bibnamefont {Reichling}}, \bibinfo
  {author} {\bibfnamefont {M.}~\bibnamefont {Abe}}, \bibinfo {author}
  {\bibfnamefont {Y.}~\bibnamefont {Sugimoto}},\ and\ \bibinfo {author}
  {\bibfnamefont {A.}~\bibnamefont {K{\"u}hnle}},\ }\bibfield  {title}
  {\enquote {\bibinfo {title} {{Flexible drift-compensation system for precise
  3D force mapping in severe drift environments}},}\ }\href
  {https://doi.org/10.1063/1.3600453} {\bibfield  {journal} {\bibinfo
  {journal} {Review of Scientific Instruments}\ }\textbf {\bibinfo {volume}
  {82}},\ \bibinfo {pages} {063704} (\bibinfo {year} {2011})}\BibitemShut
  {NoStop}%
\bibitem [{\citenamefont {Albrektsen}\ \emph {et~al.}(1994)\citenamefont
  {Albrektsen}, \citenamefont {Salemink}, \citenamefont {Mo/rch},\ and\
  \citenamefont {Th{\"o}len}}]{sharp1}%
  \BibitemOpen
  \bibfield  {author} {\bibinfo {author} {\bibfnamefont {O.}~\bibnamefont
  {Albrektsen}}, \bibinfo {author} {\bibfnamefont {H.~W.~M.}\ \bibnamefont
  {Salemink}}, \bibinfo {author} {\bibfnamefont {K.~A.}\ \bibnamefont
  {Mo/rch}},\ and\ \bibinfo {author} {\bibfnamefont {A.~R.}\ \bibnamefont
  {Th{\"o}len}},\ }\href@noop {} {\bibfield  {journal} {\bibinfo  {journal}
  {Journal of Vacuum Science \& Technology B: Microelectronics and Nanometer
  Structures Processing, Measurement, and Phenomena}\ }\textbf {\bibinfo
  {volume} {12}},\ \bibinfo {pages} {3187--3190} (\bibinfo {year}
  {1994})}\BibitemShut {NoStop}%
\bibitem [{\citenamefont {Hapala}\ \emph {et~al.}(2014)\citenamefont {Hapala},
  \citenamefont {Kichin}, \citenamefont {Wagner}, \citenamefont {Tautz},
  \citenamefont {Temirov},\ and\ \citenamefont {Jel\'{\i}nek}}]{sharp2}%
  \BibitemOpen
  \bibfield  {author} {\bibinfo {author} {\bibfnamefont {P.}~\bibnamefont
  {Hapala}}, \bibinfo {author} {\bibfnamefont {G.}~\bibnamefont {Kichin}},
  \bibinfo {author} {\bibfnamefont {C.}~\bibnamefont {Wagner}}, \bibinfo
  {author} {\bibfnamefont {F.~S.}\ \bibnamefont {Tautz}}, \bibinfo {author}
  {\bibfnamefont {R.}~\bibnamefont {Temirov}},\ and\ \bibinfo {author}
  {\bibfnamefont {P.}~\bibnamefont {Jel\'{\i}nek}},\ }\href@noop {} {\bibfield
  {journal} {\bibinfo  {journal} {Phys. Rev. B}\ }\textbf {\bibinfo {volume}
  {90}},\ \bibinfo {pages} {085421} (\bibinfo {year} {2014})}\BibitemShut
  {NoStop}%
\bibitem [{\citenamefont {MacLeod}\ \emph {et~al.}(2020)\citenamefont
  {MacLeod}, \citenamefont {Parlane}, \citenamefont {Morrissey}, \citenamefont
  {H{\"a}se}, \citenamefont {Roch}, \citenamefont {Dettelbach}, \citenamefont
  {Moreira}, \citenamefont {Yunker}, \citenamefont {Rooney}, \citenamefont
  {Deeth}, \citenamefont {Lai}, \citenamefont {Ng}, \citenamefont {Situ},
  \citenamefont {Zhang}, \citenamefont {Elliott}, \citenamefont {Haley},
  \citenamefont {Dvorak}, \citenamefont {Aspuru-Guzik}, \citenamefont {Hein},\
  and\ \citenamefont {Berlinguette}}]{selfdriving1}%
  \BibitemOpen
  \bibfield  {author} {\bibinfo {author} {\bibfnamefont {B.~P.}\ \bibnamefont
  {MacLeod}}, \bibinfo {author} {\bibfnamefont {F.~G.~L.}\ \bibnamefont
  {Parlane}}, \bibinfo {author} {\bibfnamefont {T.~D.}\ \bibnamefont
  {Morrissey}}, \bibinfo {author} {\bibfnamefont {F.}~\bibnamefont {H{\"a}se}},
  \bibinfo {author} {\bibfnamefont {L.~M.}\ \bibnamefont {Roch}}, \bibinfo
  {author} {\bibfnamefont {K.~E.}\ \bibnamefont {Dettelbach}}, \bibinfo
  {author} {\bibfnamefont {R.}~\bibnamefont {Moreira}}, \bibinfo {author}
  {\bibfnamefont {L.~P.~E.}\ \bibnamefont {Yunker}}, \bibinfo {author}
  {\bibfnamefont {M.~B.}\ \bibnamefont {Rooney}}, \bibinfo {author}
  {\bibfnamefont {J.~R.}\ \bibnamefont {Deeth}}, \bibinfo {author}
  {\bibfnamefont {V.}~\bibnamefont {Lai}}, \bibinfo {author} {\bibfnamefont
  {G.~J.}\ \bibnamefont {Ng}}, \bibinfo {author} {\bibfnamefont
  {H.}~\bibnamefont {Situ}}, \bibinfo {author} {\bibfnamefont {R.~H.}\
  \bibnamefont {Zhang}}, \bibinfo {author} {\bibfnamefont {M.~S.}\ \bibnamefont
  {Elliott}}, \bibinfo {author} {\bibfnamefont {T.~H.}\ \bibnamefont {Haley}},
  \bibinfo {author} {\bibfnamefont {D.~J.}\ \bibnamefont {Dvorak}}, \bibinfo
  {author} {\bibfnamefont {A.}~\bibnamefont {Aspuru-Guzik}}, \bibinfo {author}
  {\bibfnamefont {J.~E.}\ \bibnamefont {Hein}},\ and\ \bibinfo {author}
  {\bibfnamefont {C.~P.}\ \bibnamefont {Berlinguette}},\ }\bibfield  {title}
  {\enquote {\bibinfo {title} {Self-driving laboratory for accelerated
  discovery of thin-film materials},}\ }\href@noop {} {\bibfield  {journal}
  {\bibinfo  {journal} {Science Advances}\ }\textbf {\bibinfo {volume} {6}},\
  \bibinfo {pages} {eaaz8867} (\bibinfo {year} {2020})}\BibitemShut {NoStop}%
\bibitem [{\citenamefont {Abolhasani}\ and\ \citenamefont
  {Kumacheva}(2023)}]{selfdriving2}%
  \BibitemOpen
  \bibfield  {author} {\bibinfo {author} {\bibfnamefont {M.}~\bibnamefont
  {Abolhasani}}\ and\ \bibinfo {author} {\bibfnamefont {E.}~\bibnamefont
  {Kumacheva}},\ }\bibfield  {title} {\enquote {\bibinfo {title} {The rise of
  self-driving labs in chemical and materials sciences},}\ }\href@noop {}
  {\bibfield  {journal} {\bibinfo  {journal} {Nature Synthesis}\ }\textbf
  {\bibinfo {volume} {2}},\ \bibinfo {pages} {483--492} (\bibinfo {year}
  {2023})}\BibitemShut {NoStop}%
\bibitem [{\citenamefont {He}\ \emph {et~al.}(2016)\citenamefont {He},
  \citenamefont {Zhang}, \citenamefont {Ren},\ and\ \citenamefont
  {Sun}}]{resnet}%
  \BibitemOpen
  \bibfield  {author} {\bibinfo {author} {\bibfnamefont {K.}~\bibnamefont
  {He}}, \bibinfo {author} {\bibfnamefont {X.}~\bibnamefont {Zhang}}, \bibinfo
  {author} {\bibfnamefont {S.}~\bibnamefont {Ren}},\ and\ \bibinfo {author}
  {\bibfnamefont {J.}~\bibnamefont {Sun}},\ }\href@noop {} {\enquote {\bibinfo
  {title} {Deep residual learning for image recognition},}\ } (\bibinfo {year}
  {2016})\BibitemShut {NoStop}%
\bibitem [{\citenamefont {Liu}\ and\ \citenamefont {Deng}(2015)}]{vgg}%
  \BibitemOpen
  \bibfield  {author} {\bibinfo {author} {\bibfnamefont {S.}~\bibnamefont
  {Liu}}\ and\ \bibinfo {author} {\bibfnamefont {W.}~\bibnamefont {Deng}},\
  }\href {https://doi.org/10.1109/ACPR.2015.7486599} {} (\bibinfo {year}
  {2015})\BibitemShut {NoStop}%
\bibitem [{\citenamefont {Gordon}\ and\ \citenamefont
  {Moriarty}(2020)}]{spm_PERSPECTIVE}%
  \BibitemOpen
  \bibfield  {author} {\bibinfo {author} {\bibfnamefont {O.~M.}\ \bibnamefont
  {Gordon}}\ and\ \bibinfo {author} {\bibfnamefont {P.~J.}\ \bibnamefont
  {Moriarty}},\ }\href {https://doi.org/10.1088/2632-2153/ab7d2f} {\bibfield
  {journal} {\bibinfo  {journal} {Machine Learning: Science and Technology}\
  }\textbf {\bibinfo {volume} {1}},\ \bibinfo {pages} {023001} (\bibinfo {year}
  {2020})}\BibitemShut {NoStop}%
\bibitem [{\citenamefont {Farley}\ \emph {et~al.}(2020)\citenamefont {Farley},
  \citenamefont {Hodgkinson}, \citenamefont {Gordon}, \citenamefont {Turner},
  \citenamefont {Soltoggio}, \citenamefont {Moriarty},\ and\ \citenamefont
  {Hunsicker}}]{ai_ana1}%
  \BibitemOpen
  \bibfield  {author} {\bibinfo {author} {\bibfnamefont {S.}~\bibnamefont
  {Farley}}, \bibinfo {author} {\bibfnamefont {J.~E.~A.}\ \bibnamefont
  {Hodgkinson}}, \bibinfo {author} {\bibfnamefont {O.~M.}\ \bibnamefont
  {Gordon}}, \bibinfo {author} {\bibfnamefont {J.}~\bibnamefont {Turner}},
  \bibinfo {author} {\bibfnamefont {A.}~\bibnamefont {Soltoggio}}, \bibinfo
  {author} {\bibfnamefont {P.~J.}\ \bibnamefont {Moriarty}},\ and\ \bibinfo
  {author} {\bibfnamefont {E.}~\bibnamefont {Hunsicker}},\ }\bibfield  {title}
  {\enquote {\bibinfo {title} {Improving the segmentation of scanning probe
  microscope images using convolutional neural networks},}\ }\href
  {https://doi.org/10.1088/2632-2153/abc81c} {\bibfield  {journal} {\bibinfo
  {journal} {Machine Learning: Science and Technology}\ }\textbf {\bibinfo
  {volume} {2}},\ \bibinfo {pages} {015015} (\bibinfo {year}
  {2020})}\BibitemShut {NoStop}%
\bibitem [{\citenamefont {Gordon}\ \emph {et~al.}(2020)\citenamefont {Gordon},
  \citenamefont {Hodgkinson}, \citenamefont {Farley}, \citenamefont
  {Hunsicker},\ and\ \citenamefont {Moriarty}}]{ai_ana2}%
  \BibitemOpen
  \bibfield  {author} {\bibinfo {author} {\bibfnamefont {O.~M.}\ \bibnamefont
  {Gordon}}, \bibinfo {author} {\bibfnamefont {J.~E.~A.}\ \bibnamefont
  {Hodgkinson}}, \bibinfo {author} {\bibfnamefont {S.~M.}\ \bibnamefont
  {Farley}}, \bibinfo {author} {\bibfnamefont {E.~L.}\ \bibnamefont
  {Hunsicker}},\ and\ \bibinfo {author} {\bibfnamefont {P.~J.}\ \bibnamefont
  {Moriarty}},\ }\href@noop {} {\bibfield  {journal} {\bibinfo  {journal} {Nano
  Letters}\ }\textbf {\bibinfo {volume} {20}},\ \bibinfo {pages} {7688--7693}
  (\bibinfo {year} {2020})}\BibitemShut {NoStop}%
\bibitem [{\citenamefont {Hofer}\ \emph {et~al.}(2021)\citenamefont {Hofer},
  \citenamefont {Krstaji{\'c}}, \citenamefont {Juh{\'a}sz}, \citenamefont
  {Marchant},\ and\ \citenamefont {Smith}}]{ai_ana3}%
  \BibitemOpen
  \bibfield  {author} {\bibinfo {author} {\bibfnamefont {L.~R.}\ \bibnamefont
  {Hofer}}, \bibinfo {author} {\bibfnamefont {M.}~\bibnamefont {Krstaji{\'c}}},
  \bibinfo {author} {\bibfnamefont {P.}~\bibnamefont {Juh{\'a}sz}}, \bibinfo
  {author} {\bibfnamefont {A.~L.}\ \bibnamefont {Marchant}},\ and\ \bibinfo
  {author} {\bibfnamefont {R.~P.}\ \bibnamefont {Smith}},\ }\bibfield  {title}
  {\enquote {\bibinfo {title} {Atom cloud detection and segmentation using a
  deep neural network},}\ }\href {https://doi.org/10.1088/2632-2153/abf5ee}
  {\bibfield  {journal} {\bibinfo  {journal} {Machine Learning: Science and
  Technology}\ }\textbf {\bibinfo {volume} {2}},\ \bibinfo {pages} {045008}
  (\bibinfo {year} {2021})}\BibitemShut {NoStop}%
\bibitem [{\citenamefont {Lin}\ \emph {et~al.}(2021)\citenamefont {Lin},
  \citenamefont {Zhang}, \citenamefont {Wang}, \citenamefont {Yang},\ and\
  \citenamefont {Xin}}]{ai_ana4}%
  \BibitemOpen
  \bibfield  {author} {\bibinfo {author} {\bibfnamefont {R.}~\bibnamefont
  {Lin}}, \bibinfo {author} {\bibfnamefont {R.}~\bibnamefont {Zhang}}, \bibinfo
  {author} {\bibfnamefont {C.}~\bibnamefont {Wang}}, \bibinfo {author}
  {\bibfnamefont {X.-Q.}\ \bibnamefont {Yang}},\ and\ \bibinfo {author}
  {\bibfnamefont {H.~L.}\ \bibnamefont {Xin}},\ }\href
  {https://doi.org/10.1038/s41598-021-84499-w} {\bibfield  {journal} {\bibinfo
  {journal} {Scientific Reports}\ }\textbf {\bibinfo {volume} {11}},\ \bibinfo
  {pages} {5386} (\bibinfo {year} {2021})}\BibitemShut {NoStop}%
\bibitem [{\citenamefont {Yang}\ \emph {et~al.}(2021)\citenamefont {Yang},
  \citenamefont {Choi}, \citenamefont {Cho}, \citenamefont {Agyapong-Fordjour},
  \citenamefont {Park}, \citenamefont {Yun}, \citenamefont {Kim}, \citenamefont
  {Han}, \citenamefont {Lee}, \citenamefont {Kim},\ and\ \citenamefont
  {Kim}}]{ai_ana5}%
  \BibitemOpen
  \bibfield  {author} {\bibinfo {author} {\bibfnamefont {S.-H.}\ \bibnamefont
  {Yang}}, \bibinfo {author} {\bibfnamefont {W.}~\bibnamefont {Choi}}, \bibinfo
  {author} {\bibfnamefont {B.~W.}\ \bibnamefont {Cho}}, \bibinfo {author}
  {\bibfnamefont {F.~O.-T.}\ \bibnamefont {Agyapong-Fordjour}}, \bibinfo
  {author} {\bibfnamefont {S.}~\bibnamefont {Park}}, \bibinfo {author}
  {\bibfnamefont {S.~J.}\ \bibnamefont {Yun}}, \bibinfo {author} {\bibfnamefont
  {H.-J.}\ \bibnamefont {Kim}}, \bibinfo {author} {\bibfnamefont {Y.-K.}\
  \bibnamefont {Han}}, \bibinfo {author} {\bibfnamefont {Y.~H.}\ \bibnamefont
  {Lee}}, \bibinfo {author} {\bibfnamefont {K.~K.}\ \bibnamefont {Kim}},\ and\
  \bibinfo {author} {\bibfnamefont {Y.-M.}\ \bibnamefont {Kim}},\ }\href
  {https://doi.org/https://doi.org/10.1002/advs.202101099} {\bibfield
  {journal} {\bibinfo  {journal} {Advanced Science}\ }\textbf {\bibinfo
  {volume} {8}},\ \bibinfo {pages} {2101099} (\bibinfo {year}
  {2021})}\BibitemShut {NoStop}%
\bibitem [{\citenamefont {Wang}\ \emph {et~al.}(2021)\citenamefont {Wang},
  \citenamefont {Zhu}, \citenamefont {Blackwell},\ and\ \citenamefont
  {Fischer}}]{ai_sts}%
  \BibitemOpen
  \bibfield  {author} {\bibinfo {author} {\bibfnamefont {S.}~\bibnamefont
  {Wang}}, \bibinfo {author} {\bibfnamefont {J.}~\bibnamefont {Zhu}}, \bibinfo
  {author} {\bibfnamefont {R.}~\bibnamefont {Blackwell}},\ and\ \bibinfo
  {author} {\bibfnamefont {F.~R.}\ \bibnamefont {Fischer}},\ }\bibfield
  {title} {\enquote {\bibinfo {title} {Automated tip conditioning for scanning
  tunneling spectroscopy},}\ }\href {https://doi.org/10.1021/acs.jpca.0c10731}
  {\bibfield  {journal} {\bibinfo  {journal} {The Journal of Physical Chemistry
  A}\ }\textbf {\bibinfo {volume} {125}},\ \bibinfo {pages} {1384--1390}
  (\bibinfo {year} {2021})},\ \bibinfo {note} {pMID: 33560124},\ \Eprint
  {https://arxiv.org/abs/https://doi.org/10.1021/acs.jpca.0c10731}
  {https://doi.org/10.1021/acs.jpca.0c10731} \BibitemShut {NoStop}%
\bibitem [{\citenamefont {Krull}\ \emph {et~al.}(2020)\citenamefont {Krull},
  \citenamefont {Hirsch}, \citenamefont {Rother}, \citenamefont {Schiffrin},\
  and\ \citenamefont {Krull}}]{ai_scan}%
  \BibitemOpen
  \bibfield  {author} {\bibinfo {author} {\bibfnamefont {A.}~\bibnamefont
  {Krull}}, \bibinfo {author} {\bibfnamefont {P.}~\bibnamefont {Hirsch}},
  \bibinfo {author} {\bibfnamefont {C.}~\bibnamefont {Rother}}, \bibinfo
  {author} {\bibfnamefont {A.}~\bibnamefont {Schiffrin}},\ and\ \bibinfo
  {author} {\bibfnamefont {C.}~\bibnamefont {Krull}},\ }\href
  {https://doi.org/10.1038/s42005-020-0317-3} {\bibfield  {journal} {\bibinfo
  {journal} {Communications Physics}\ }\textbf {\bibinfo {volume} {3}},\
  \bibinfo {pages} {54} (\bibinfo {year} {2020})}\BibitemShut {NoStop}%
\bibitem [{\citenamefont {Thomas}\ \emph {et~al.}(2022)\citenamefont {Thomas},
  \citenamefont {Rossi}, \citenamefont {Smalley}, \citenamefont {Francaviglia},
  \citenamefont {Yu}, \citenamefont {Zhang}, \citenamefont {Kumari},
  \citenamefont {Robinson}, \citenamefont {Terrones}, \citenamefont {Ishigami},
  \citenamefont {Rotenberg}, \citenamefont {Barnard}, \citenamefont {Raja},
  \citenamefont {Wong}, \citenamefont {Ogletree}, \citenamefont {Noack},\ and\
  \citenamefont {Weber-Bargioni}}]{ai_scan2}%
  \BibitemOpen
  \bibfield  {author} {\bibinfo {author} {\bibfnamefont {J.~C.}\ \bibnamefont
  {Thomas}}, \bibinfo {author} {\bibfnamefont {A.}~\bibnamefont {Rossi}},
  \bibinfo {author} {\bibfnamefont {D.}~\bibnamefont {Smalley}}, \bibinfo
  {author} {\bibfnamefont {L.}~\bibnamefont {Francaviglia}}, \bibinfo {author}
  {\bibfnamefont {Z.}~\bibnamefont {Yu}}, \bibinfo {author} {\bibfnamefont
  {T.}~\bibnamefont {Zhang}}, \bibinfo {author} {\bibfnamefont
  {S.}~\bibnamefont {Kumari}}, \bibinfo {author} {\bibfnamefont {J.~A.}\
  \bibnamefont {Robinson}}, \bibinfo {author} {\bibfnamefont {M.}~\bibnamefont
  {Terrones}}, \bibinfo {author} {\bibfnamefont {M.}~\bibnamefont {Ishigami}},
  \bibinfo {author} {\bibfnamefont {E.}~\bibnamefont {Rotenberg}}, \bibinfo
  {author} {\bibfnamefont {E.~S.}\ \bibnamefont {Barnard}}, \bibinfo {author}
  {\bibfnamefont {A.}~\bibnamefont {Raja}}, \bibinfo {author} {\bibfnamefont
  {E.}~\bibnamefont {Wong}}, \bibinfo {author} {\bibfnamefont {D.~F.}\
  \bibnamefont {Ogletree}}, \bibinfo {author} {\bibfnamefont {M.~M.}\
  \bibnamefont {Noack}},\ and\ \bibinfo {author} {\bibfnamefont
  {A.}~\bibnamefont {Weber-Bargioni}},\ }\href@noop {} {\bibfield  {journal}
  {\bibinfo  {journal} {npj Computational Materials}\ }\textbf {\bibinfo
  {volume} {8}},\ \bibinfo {pages} {99} (\bibinfo {year} {2022})}\BibitemShut
  {NoStop}%
\bibitem [{\citenamefont {Chen}\ \emph {et~al.}(2022)\citenamefont {Chen},
  \citenamefont {Aapro}, \citenamefont {Kipnis}, \citenamefont {Ilin},
  \citenamefont {Liljeroth},\ and\ \citenamefont {Foster}}]{ai_am}%
  \BibitemOpen
  \bibfield  {author} {\bibinfo {author} {\bibfnamefont {I.-J.}\ \bibnamefont
  {Chen}}, \bibinfo {author} {\bibfnamefont {M.}~\bibnamefont {Aapro}},
  \bibinfo {author} {\bibfnamefont {A.}~\bibnamefont {Kipnis}}, \bibinfo
  {author} {\bibfnamefont {A.}~\bibnamefont {Ilin}}, \bibinfo {author}
  {\bibfnamefont {P.}~\bibnamefont {Liljeroth}},\ and\ \bibinfo {author}
  {\bibfnamefont {A.~S.}\ \bibnamefont {Foster}},\ }\bibfield  {title}
  {\enquote {\bibinfo {title} {Precise atom manipulation through deep
  reinforcement learning},}\ }\href
  {https://doi.org/10.1038/s41467-022-35149-w} {\bibfield  {journal} {\bibinfo
  {journal} {Nature Communications}\ }\textbf {\bibinfo {volume} {13}},\
  \bibinfo {pages} {7499} (\bibinfo {year} {2022})}\BibitemShut {NoStop}%
\bibitem [{\citenamefont {Diao}, \citenamefont {Hou},\ and\ \citenamefont
  {Abe}(2023)}]{tipfix}%
  \BibitemOpen
  \bibfield  {author} {\bibinfo {author} {\bibfnamefont {Z.}~\bibnamefont
  {Diao}}, \bibinfo {author} {\bibfnamefont {L.}~\bibnamefont {Hou}},\ and\
  \bibinfo {author} {\bibfnamefont {M.}~\bibnamefont {Abe}},\ }\bibfield
  {title} {\enquote {\bibinfo {title} {Probe conditioning via convolution
  neural network for scanning probe microscopy automation},}\ }\href
  {https://doi.org/10.35848/1882-0786/acecd6} {\bibfield  {journal} {\bibinfo
  {journal} {Applied Physics Express}\ }\textbf {\bibinfo {volume} {16}},\
  \bibinfo {pages} {085002} (\bibinfo {year} {2023})}\BibitemShut {NoStop}%
\bibitem [{\citenamefont {Sugimoto}\ \emph {et~al.}(2013)\citenamefont
  {Sugimoto}, \citenamefont {Yurtsever}, \citenamefont {Abe}, \citenamefont
  {Morita}, \citenamefont {Ondr\'{a}\v{c}ek}, \citenamefont {Pou},
  \citenamefont {P\'{e}rez},\ and\ \citenamefont
  {Jel\'{i}nek}}]{doi:10.1021/nn403097p}%
  \BibitemOpen
  \bibfield  {author} {\bibinfo {author} {\bibfnamefont {Y.}~\bibnamefont
  {Sugimoto}}, \bibinfo {author} {\bibfnamefont {A.}~\bibnamefont {Yurtsever}},
  \bibinfo {author} {\bibfnamefont {M.}~\bibnamefont {Abe}}, \bibinfo {author}
  {\bibfnamefont {S.}~\bibnamefont {Morita}}, \bibinfo {author} {\bibfnamefont
  {M.}~\bibnamefont {Ondr\'{a}\v{c}ek}}, \bibinfo {author} {\bibfnamefont
  {P.}~\bibnamefont {Pou}}, \bibinfo {author} {\bibfnamefont {R.}~\bibnamefont
  {P\'{e}rez}},\ and\ \bibinfo {author} {\bibfnamefont {P.}~\bibnamefont
  {Jel\'{i}nek}},\ }\bibfield  {title} {\enquote {\bibinfo {title} {Role of tip
  chemical reactivity on atom manipulation process in dynamic force
  microscopy},}\ }\href {https://doi.org/10.1021/nn403097p} {\bibfield
  {journal} {\bibinfo  {journal} {ACS Nano}\ }\textbf {\bibinfo {volume} {7}},\
  \bibinfo {pages} {7370--7376} (\bibinfo {year} {2013})},\ \bibinfo {note}
  {pMID: 23906095},\ \Eprint
  {https://arxiv.org/abs/https://doi.org/10.1021/nn403097p}
  {https://doi.org/10.1021/nn403097p} \BibitemShut {NoStop}%
\bibitem [{\citenamefont {Sugimoto}\ \emph
  {et~al.}(2008{\natexlab{b}})\citenamefont {Sugimoto}, \citenamefont {Miki},
  \citenamefont {Abe},\ and\ \citenamefont {Morita}}]{PhysRevB.78.205305}%
  \BibitemOpen
  \bibfield  {author} {\bibinfo {author} {\bibfnamefont {Y.}~\bibnamefont
  {Sugimoto}}, \bibinfo {author} {\bibfnamefont {K.}~\bibnamefont {Miki}},
  \bibinfo {author} {\bibfnamefont {M.}~\bibnamefont {Abe}},\ and\ \bibinfo
  {author} {\bibfnamefont {S.}~\bibnamefont {Morita}},\ }\bibfield  {title}
  {\enquote {\bibinfo {title} {Statistics of lateral atom manipulation by
  atomic force microscopy at room temperature},}\ }\href
  {https://doi.org/10.1103/PhysRevB.78.205305} {\bibfield  {journal} {\bibinfo
  {journal} {Phys. Rev. B}\ }\textbf {\bibinfo {volume} {78}},\ \bibinfo
  {pages} {205305} (\bibinfo {year} {2008}{\natexlab{b}})}\BibitemShut
  {NoStop}%
\bibitem [{\citenamefont {Ming}\ \emph {et~al.}(2011)\citenamefont {Ming},
  \citenamefont {Wang}, \citenamefont {Pan}, \citenamefont {Liu}, \citenamefont
  {Zhang}, \citenamefont {Yang},\ and\ \citenamefont
  {Xiao}}]{doi:10.1021/nn202636g}%
  \BibitemOpen
  \bibfield  {author} {\bibinfo {author} {\bibfnamefont {F.}~\bibnamefont
  {Ming}}, \bibinfo {author} {\bibfnamefont {K.}~\bibnamefont {Wang}}, \bibinfo
  {author} {\bibfnamefont {S.}~\bibnamefont {Pan}}, \bibinfo {author}
  {\bibfnamefont {J.}~\bibnamefont {Liu}}, \bibinfo {author} {\bibfnamefont
  {X.}~\bibnamefont {Zhang}}, \bibinfo {author} {\bibfnamefont
  {J.}~\bibnamefont {Yang}},\ and\ \bibinfo {author} {\bibfnamefont
  {X.}~\bibnamefont {Xiao}},\ }\bibfield  {title} {\enquote {\bibinfo {title}
  {Assembling and disassembling ag clusters on si(111)-(7×7) by vertical
  atomic manipulation},}\ }\href {https://doi.org/10.1021/nn202636g} {\bibfield
   {journal} {\bibinfo  {journal} {ACS Nano}\ }\textbf {\bibinfo {volume}
  {5}},\ \bibinfo {pages} {7608--7616} (\bibinfo {year} {2011})},\ \bibinfo
  {note} {pMID: 21819127}\BibitemShut {NoStop}%
\bibitem [{\citenamefont {Sugimoto}\ \emph {et~al.}(2014)\citenamefont
  {Sugimoto}, \citenamefont {Yurtsever}, \citenamefont {Hirayama},
  \citenamefont {Abe},\ and\ \citenamefont {Morita}}]{sugimoto2014NatCom}%
  \BibitemOpen
  \bibfield  {author} {\bibinfo {author} {\bibfnamefont {Y.}~\bibnamefont
  {Sugimoto}}, \bibinfo {author} {\bibfnamefont {A.}~\bibnamefont {Yurtsever}},
  \bibinfo {author} {\bibfnamefont {N.}~\bibnamefont {Hirayama}}, \bibinfo
  {author} {\bibfnamefont {M.}~\bibnamefont {Abe}},\ and\ \bibinfo {author}
  {\bibfnamefont {S.}~\bibnamefont {Morita}},\ }\bibfield  {title} {\enquote
  {\bibinfo {title} {Mechanical gate control for atom-by-atom cluster assembly
  with scanning probe microscopy},}\ }\href
  {https://doi.org/10.1038/ncomms5360} {\bibfield  {journal} {\bibinfo
  {journal} {Nature Communications}\ }\textbf {\bibinfo {volume} {5}},\
  \bibinfo {pages} {4360} (\bibinfo {year} {2014})}\BibitemShut {NoStop}%
\bibitem [{\citenamefont {Inami}\ \emph {et~al.}(2015)\citenamefont {Inami},
  \citenamefont {Hamada}, \citenamefont {Ueda}, \citenamefont {Abe},
  \citenamefont {Morita},\ and\ \citenamefont {Sugimoto}}]{Inami2015NatCom}%
  \BibitemOpen
  \bibfield  {author} {\bibinfo {author} {\bibfnamefont {E.}~\bibnamefont
  {Inami}}, \bibinfo {author} {\bibfnamefont {I.}~\bibnamefont {Hamada}},
  \bibinfo {author} {\bibfnamefont {K.}~\bibnamefont {Ueda}}, \bibinfo {author}
  {\bibfnamefont {M.}~\bibnamefont {Abe}}, \bibinfo {author} {\bibfnamefont
  {S.}~\bibnamefont {Morita}},\ and\ \bibinfo {author} {\bibfnamefont
  {Y.}~\bibnamefont {Sugimoto}},\ }\bibfield  {title} {\enquote {\bibinfo
  {title} {Room-temperature-concerted switch made of a binary atom cluster},}\
  }\href {https://doi.org/10.1038/ncomms7231} {\bibfield  {journal} {\bibinfo
  {journal} {Nature Communications}\ }\textbf {\bibinfo {volume} {6}},\
  \bibinfo {pages} {6231} (\bibinfo {year} {2015})}\BibitemShut {NoStop}%
\bibitem [{\citenamefont {Hwang}, \citenamefont {Ho},\ and\ \citenamefont
  {Tsong}(1999)}]{Hwang1999DYNAMICBO}%
  \BibitemOpen
  \bibfield  {author} {\bibinfo {author} {\bibfnamefont {I.-S.}\ \bibnamefont
  {Hwang}}, \bibinfo {author} {\bibfnamefont {M.-S.}\ \bibnamefont {Ho}},\ and\
  \bibinfo {author} {\bibfnamefont {T.~T.}\ \bibnamefont {Tsong}},\ }\bibfield
  {title} {\enquote {\bibinfo {title} {Dynamic behavior of si magic clusters on
  si(111) surfaces},}\ }\href
  {https://api.semanticscholar.org/CorpusID:122162625} {\bibfield  {journal}
  {\bibinfo  {journal} {Physical Review Letters}\ }\textbf {\bibinfo {volume}
  {83}},\ \bibinfo {pages} {120--123} (\bibinfo {year} {1999})}\BibitemShut
  {NoStop}%
\bibitem [{\citenamefont {Zhang}\ \emph {et~al.}(2005)\citenamefont {Zhang},
  \citenamefont {Chen}, \citenamefont {Wang}, \citenamefont {Yang},
  \citenamefont {Su}, \citenamefont {Chan}, \citenamefont {Loy},\ and\
  \citenamefont {Xiao}}]{PhysRevLett.94.176104}%
  \BibitemOpen
  \bibfield  {author} {\bibinfo {author} {\bibfnamefont {C.}~\bibnamefont
  {Zhang}}, \bibinfo {author} {\bibfnamefont {G.}~\bibnamefont {Chen}},
  \bibinfo {author} {\bibfnamefont {K.}~\bibnamefont {Wang}}, \bibinfo {author}
  {\bibfnamefont {H.}~\bibnamefont {Yang}}, \bibinfo {author} {\bibfnamefont
  {T.}~\bibnamefont {Su}}, \bibinfo {author} {\bibfnamefont {C.~T.}\
  \bibnamefont {Chan}}, \bibinfo {author} {\bibfnamefont {M.~M.~T.}\
  \bibnamefont {Loy}},\ and\ \bibinfo {author} {\bibfnamefont {X.}~\bibnamefont
  {Xiao}},\ }\bibfield  {title} {\enquote {\bibinfo {title} {Experimental and
  theoretical investigation of single cu, ag, and au atoms adsorbed on
  $\mathrm{Si}(111)\mathrm{\text{\ensuremath{-}}}(7\ifmmode\times\else\texttimes\fi{}7)$},}\
  }\href {https://doi.org/10.1103/PhysRevLett.94.176104} {\bibfield  {journal}
  {\bibinfo  {journal} {Phys. Rev. Lett.}\ }\textbf {\bibinfo {volume} {94}},\
  \bibinfo {pages} {176104} (\bibinfo {year} {2005})}\BibitemShut {NoStop}%
\bibitem [{\citenamefont {O\ifmmode~\check{s}\else \v{s}\fi{}t'\'adal}\ \emph
  {et~al.}(2005)\citenamefont {O\ifmmode~\check{s}\else \v{s}\fi{}t'\'adal},
  \citenamefont {Koc\'an}, \citenamefont {Sobot\'{\i}k},\ and\ \citenamefont
  {Pudl}}]{PhysRevLett.95.146101}%
  \BibitemOpen
  \bibfield  {author} {\bibinfo {author} {\bibfnamefont {I.}~\bibnamefont
  {O\ifmmode~\check{s}\else \v{s}\fi{}t'\'adal}}, \bibinfo {author}
  {\bibfnamefont {P.}~\bibnamefont {Koc\'an}}, \bibinfo {author} {\bibfnamefont
  {P.}~\bibnamefont {Sobot\'{\i}k}},\ and\ \bibinfo {author} {\bibfnamefont
  {J.}~\bibnamefont {Pudl}},\ }\bibfield  {title} {\enquote {\bibinfo {title}
  {Direct observation of long-range assisted formation of ag clusters on
  $\mathrm{Si}(111)7\ifmmode\times\else\texttimes\fi{}7$},}\ }\href
  {https://doi.org/10.1103/PhysRevLett.95.146101} {\bibfield  {journal}
  {\bibinfo  {journal} {Phys. Rev. Lett.}\ }\textbf {\bibinfo {volume} {95}},\
  \bibinfo {pages} {146101} (\bibinfo {year} {2005})}\BibitemShut {NoStop}%
\bibitem [{\citenamefont {Wang}\ \emph {et~al.}(2008)\citenamefont {Wang},
  \citenamefont {Chen}, \citenamefont {Zhang}, \citenamefont {Loy},\ and\
  \citenamefont {Xiao}}]{PhysRevLett.101.266107}%
  \BibitemOpen
  \bibfield  {author} {\bibinfo {author} {\bibfnamefont {K.}~\bibnamefont
  {Wang}}, \bibinfo {author} {\bibfnamefont {G.}~\bibnamefont {Chen}}, \bibinfo
  {author} {\bibfnamefont {C.}~\bibnamefont {Zhang}}, \bibinfo {author}
  {\bibfnamefont {M.~M.~T.}\ \bibnamefont {Loy}},\ and\ \bibinfo {author}
  {\bibfnamefont {X.}~\bibnamefont {Xiao}},\ }\bibfield  {title} {\enquote
  {\bibinfo {title} {Intermixing of intrabasin and interbasin diffusion of a
  single ag atom on
  $\mathrm{Si}(111)\mathrm{\text{\ensuremath{-}}}(7\ifmmode\times\else\texttimes\fi{}7)$},}\
  }\href {https://doi.org/10.1103/PhysRevLett.101.266107} {\bibfield  {journal}
  {\bibinfo  {journal} {Phys. Rev. Lett.}\ }\textbf {\bibinfo {volume} {101}},\
  \bibinfo {pages} {266107} (\bibinfo {year} {2008})}\BibitemShut {NoStop}%
\bibitem [{\citenamefont {Osiecki}, \citenamefont {Suto},\ and\ \citenamefont
  {Chutia}(2022)}]{Osiecki2022NatCom}%
  \BibitemOpen
  \bibfield  {author} {\bibinfo {author} {\bibfnamefont {J.~R.}\ \bibnamefont
  {Osiecki}}, \bibinfo {author} {\bibfnamefont {S.}~\bibnamefont {Suto}},\ and\
  \bibinfo {author} {\bibfnamefont {A.}~\bibnamefont {Chutia}},\ }\bibfield
  {title} {\enquote {\bibinfo {title} {Periodic corner holes on the
  si(111)-7×7 surface can trap silver atoms},}\ }\href
  {https://doi.org/10.1038/s41467-022-29768-6} {\bibfield  {journal} {\bibinfo
  {journal} {Nature Communications}\ }\textbf {\bibinfo {volume} {13}},\
  \bibinfo {pages} {2973} (\bibinfo {year} {2022})}\BibitemShut {NoStop}%
\bibitem [{\citenamefont {Custance}\ \emph
  {et~al.}(2001{\natexlab{b}})\citenamefont {Custance}, \citenamefont
  {Brihuega}, \citenamefont {G\'{o}mez-Rodr\'{i}guez},\ and\ \citenamefont
  {Bar\'{o}}}]{CUSTANCE20011406}%
  \BibitemOpen
  \bibfield  {author} {\bibinfo {author} {\bibfnamefont {O.}~\bibnamefont
  {Custance}}, \bibinfo {author} {\bibfnamefont {I.}~\bibnamefont {Brihuega}},
  \bibinfo {author} {\bibfnamefont {J.}~\bibnamefont
  {G\'{o}mez-Rodr\'{i}guez}},\ and\ \bibinfo {author} {\bibfnamefont
  {A.}~\bibnamefont {Bar\'{o}}},\ }\bibfield  {title} {\enquote {\bibinfo
  {title} {Initial stages of sn adsorption on si(111)-(7×7)},}\ }\href
  {https://doi.org/https://doi.org/10.1016/S0039-6028(01)00732-4} {\bibfield
  {journal} {\bibinfo  {journal} {Surface Science}\ }\textbf {\bibinfo {volume}
  {482-485}},\ \bibinfo {pages} {1406--1412} (\bibinfo {year}
  {2001}{\natexlab{b}})}\BibitemShut {NoStop}%
\bibitem [{\citenamefont {Diao}\ \emph {et~al.}(2023)\citenamefont {Diao},
  \citenamefont {Ueda}, \citenamefont {Hou}, \citenamefont {Yamashita},
  \citenamefont {Custance},\ and\ \citenamefont {Abe}}]{drift}%
  \BibitemOpen
  \bibfield  {author} {\bibinfo {author} {\bibfnamefont {Z.}~\bibnamefont
  {Diao}}, \bibinfo {author} {\bibfnamefont {K.}~\bibnamefont {Ueda}}, \bibinfo
  {author} {\bibfnamefont {L.}~\bibnamefont {Hou}}, \bibinfo {author}
  {\bibfnamefont {H.}~\bibnamefont {Yamashita}}, \bibinfo {author}
  {\bibfnamefont {O.}~\bibnamefont {Custance}},\ and\ \bibinfo {author}
  {\bibfnamefont {M.}~\bibnamefont {Abe}},\ }\bibfield  {title} {\enquote
  {\bibinfo {title} {{Automatic drift compensation for nanoscale imaging using
  feature point matching}},}\ }\href {https://doi.org/10.1063/5.0139330}
  {\bibfield  {journal} {\bibinfo  {journal} {Applied Physics Letters}\
  }\textbf {\bibinfo {volume} {122}} (\bibinfo {year} {2023}),\
  10.1063/5.0139330},\ \bibinfo {note} {121601}\BibitemShut {NoStop}%
\bibitem [{\citenamefont {Hamers}, \citenamefont {Tromp},\ and\ \citenamefont
  {Demuth}(1986)}]{si_sts1}%
  \BibitemOpen
  \bibfield  {author} {\bibinfo {author} {\bibfnamefont {R.~J.}\ \bibnamefont
  {Hamers}}, \bibinfo {author} {\bibfnamefont {R.~M.}\ \bibnamefont {Tromp}},\
  and\ \bibinfo {author} {\bibfnamefont {J.~E.}\ \bibnamefont {Demuth}},\
  }\bibfield  {title} {\enquote {\bibinfo {title} {Surface electronic structure
  of si (111)-(7\ifmmode\times\else\texttimes\fi{}7) resolved in real space},}\
  }\href@noop {} {\bibfield  {journal} {\bibinfo  {journal} {Phys. Rev. Lett.}\
  }\textbf {\bibinfo {volume} {56}},\ \bibinfo {pages} {1972--1975} (\bibinfo
  {year} {1986})}\BibitemShut {NoStop}%
\bibitem [{\citenamefont {Neddermeyer}\ and\ \citenamefont
  {Tosch}(1988)}]{si_sts2}%
  \BibitemOpen
  \bibfield  {author} {\bibinfo {author} {\bibfnamefont {H.}~\bibnamefont
  {Neddermeyer}}\ and\ \bibinfo {author} {\bibfnamefont {S.}~\bibnamefont
  {Tosch}},\ }\bibfield  {title} {\enquote {\bibinfo {title} {Scanning
  tunneling spectroscopy on si},}\ }\href@noop {} {\bibfield  {journal}
  {\bibinfo  {journal} {Ultramicroscopy}\ }\textbf {\bibinfo {volume} {25}},\
  \bibinfo {pages} {135--147} (\bibinfo {year} {1988})}\BibitemShut {NoStop}%
\bibitem [{\citenamefont {Avouris}\ and\ \citenamefont
  {Wolkow}(1989)}]{si_sts3}%
  \BibitemOpen
  \bibfield  {author} {\bibinfo {author} {\bibfnamefont {P.}~\bibnamefont
  {Avouris}}\ and\ \bibinfo {author} {\bibfnamefont {R.}~\bibnamefont
  {Wolkow}},\ }\bibfield  {title} {\enquote {\bibinfo {title} {Atom-resolved
  surface chemistry studied by scanning tunneling microscopy and
  spectroscopy},}\ }\href@noop {} {\bibfield  {journal} {\bibinfo  {journal}
  {Phys. Rev. B}\ }\textbf {\bibinfo {volume} {39}},\ \bibinfo {pages}
  {5091--5100} (\bibinfo {year} {1989})}\BibitemShut {NoStop}%
\bibitem [{\citenamefont {Myslive\ifmmode~\check{c}\else \v{c}\fi{}ek}\ \emph
  {et~al.}(2006)\citenamefont {Myslive\ifmmode~\check{c}\else \v{c}\fi{}ek},
  \citenamefont {Str\'o\ifmmode~\dot{z}\else \.{z}\fi{}ecka}, \citenamefont
  {Steffl}, \citenamefont {Sobot\'{\i}k}, \citenamefont
  {O\ifmmode~\check{s}\else \v{s}\fi{}t'\'adal},\ and\ \citenamefont
  {Voigtl\"ander}}]{PhysRevB.73.161302}%
  \BibitemOpen
  \bibfield  {author} {\bibinfo {author} {\bibfnamefont {J.}~\bibnamefont
  {Myslive\ifmmode~\check{c}\else \v{c}\fi{}ek}}, \bibinfo {author}
  {\bibfnamefont {A.}~\bibnamefont {Str\'o\ifmmode~\dot{z}\else
  \.{z}\fi{}ecka}}, \bibinfo {author} {\bibfnamefont {J.}~\bibnamefont
  {Steffl}}, \bibinfo {author} {\bibfnamefont {P.}~\bibnamefont
  {Sobot\'{\i}k}}, \bibinfo {author} {\bibfnamefont {I.}~\bibnamefont
  {O\ifmmode~\check{s}\else \v{s}\fi{}t'\'adal}},\ and\ \bibinfo {author}
  {\bibfnamefont {B.}~\bibnamefont {Voigtl\"ander}},\ }\href@noop {} {\bibfield
   {journal} {\bibinfo  {journal} {Phys. Rev. B}\ }\textbf {\bibinfo {volume}
  {73}},\ \bibinfo {pages} {161302} (\bibinfo {year} {2006})}\BibitemShut
  {NoStop}%
\bibitem [{\citenamefont {Odobescu}\ and\ \citenamefont
  {Zaitsev-Zotov}(2012)}]{Odobescu_2012}%
  \BibitemOpen
  \bibfield  {author} {\bibinfo {author} {\bibfnamefont {A.~B.}\ \bibnamefont
  {Odobescu}}\ and\ \bibinfo {author} {\bibfnamefont {S.~V.}\ \bibnamefont
  {Zaitsev-Zotov}},\ }\bibfield  {title} {\enquote {\bibinfo {title} {Energy
  gap revealed by low-temperature scanning--tunnelling spectroscopy of the
  si(111)-7×7 surface in illuminated slightly doped crystals},}\ }\href@noop
  {} {\bibfield  {journal} {\bibinfo  {journal} {Journal of Physics: Condensed
  Matter}\ }\textbf {\bibinfo {volume} {24}},\ \bibinfo {pages} {395003}
  (\bibinfo {year} {2012})}\BibitemShut {NoStop}%
\bibitem [{\citenamefont {Steinier}, \citenamefont {Termonia},\ and\
  \citenamefont {Deltour}(1972)}]{savgol}%
  \BibitemOpen
  \bibfield  {author} {\bibinfo {author} {\bibfnamefont {J.}~\bibnamefont
  {Steinier}}, \bibinfo {author} {\bibfnamefont {Y.}~\bibnamefont {Termonia}},\
  and\ \bibinfo {author} {\bibfnamefont {J.}~\bibnamefont {Deltour}},\
  }\bibfield  {title} {\enquote {\bibinfo {title} {Smoothing and
  differentiation of data by simplified least square procedure},}\ }\href
  {https://doi.org/10.1021/ac60319a045} {\bibfield  {journal} {\bibinfo
  {journal} {Analytical Chemistry}\ }\textbf {\bibinfo {volume} {44}},\
  \bibinfo {pages} {1906--1909} (\bibinfo {year} {1972})}\BibitemShut {NoStop}%
\bibitem [{\citenamefont {Tromp}, \citenamefont {Hamers},\ and\ \citenamefont
  {Demuth}(1986)}]{PhysRevB.34.1388}%
  \BibitemOpen
  \bibfield  {author} {\bibinfo {author} {\bibfnamefont {R.~M.}\ \bibnamefont
  {Tromp}}, \bibinfo {author} {\bibfnamefont {R.~J.}\ \bibnamefont {Hamers}},\
  and\ \bibinfo {author} {\bibfnamefont {J.~E.}\ \bibnamefont {Demuth}},\
  }\href {https://doi.org/10.1103/PhysRevB.34.1388} {\bibfield  {journal}
  {\bibinfo  {journal} {Phys. Rev. B}\ }\textbf {\bibinfo {volume} {34}},\
  \bibinfo {pages} {1388--1391} (\bibinfo {year} {1986})}\BibitemShut {NoStop}%
\bibitem [{\citenamefont {Himpsel}\ and\ \citenamefont
  {Fauster}(1984)}]{si_ups}%
  \BibitemOpen
  \bibfield  {author} {\bibinfo {author} {\bibfnamefont {F.~J.}\ \bibnamefont
  {Himpsel}}\ and\ \bibinfo {author} {\bibfnamefont {T.}~\bibnamefont
  {Fauster}},\ }\bibfield  {title} {\enquote {\bibinfo {title} {Probing valence
  states with photoemission and inverse photoemission},}\ }\href@noop {}
  {\bibfield  {journal} {\bibinfo  {journal} {Journal of Vacuum Science \&
  Technology A}\ }\textbf {\bibinfo {volume} {2}},\ \bibinfo {pages} {815--821}
  (\bibinfo {year} {1984})}\BibitemShut {NoStop}%
\bibitem [{\citenamefont {Himpsel}, \citenamefont {Fauster},\ and\
  \citenamefont {Hollinger}(1983)}]{si_ips}%
  \BibitemOpen
  \bibfield  {author} {\bibinfo {author} {\bibfnamefont {F.~J.}\ \bibnamefont
  {Himpsel}}, \bibinfo {author} {\bibfnamefont {T.}~\bibnamefont {Fauster}},\
  and\ \bibinfo {author} {\bibfnamefont {G.}~\bibnamefont {Hollinger}},\
  }\bibfield  {title} {\enquote {\bibinfo {title} {Electronic structure of
  si(111) surfaces},}\ }\href@noop {} {\bibfield  {journal} {\bibinfo
  {journal} {Surface Science}\ }\textbf {\bibinfo {volume} {132}},\ \bibinfo
  {pages} {22--30} (\bibinfo {year} {1983})}\BibitemShut {NoStop}%
\bibitem [{\citenamefont {Dosovitskiy}\ \emph {et~al.}(2021)\citenamefont
  {Dosovitskiy}, \citenamefont {Beyer}, \citenamefont {Kolesnikov},
  \citenamefont {Weissenborn}, \citenamefont {Zhai}, \citenamefont
  {Unterthiner}, \citenamefont {Dehghani}, \citenamefont {Minderer},
  \citenamefont {Heigold}, \citenamefont {Gelly}, \citenamefont {Uszkoreit},\
  and\ \citenamefont {Houlsby}}]{dosovitskiy2021image}%
  \BibitemOpen
  \bibfield  {author} {\bibinfo {author} {\bibfnamefont {A.}~\bibnamefont
  {Dosovitskiy}}, \bibinfo {author} {\bibfnamefont {L.}~\bibnamefont {Beyer}},
  \bibinfo {author} {\bibfnamefont {A.}~\bibnamefont {Kolesnikov}}, \bibinfo
  {author} {\bibfnamefont {D.}~\bibnamefont {Weissenborn}}, \bibinfo {author}
  {\bibfnamefont {X.}~\bibnamefont {Zhai}}, \bibinfo {author} {\bibfnamefont
  {T.}~\bibnamefont {Unterthiner}}, \bibinfo {author} {\bibfnamefont
  {M.}~\bibnamefont {Dehghani}}, \bibinfo {author} {\bibfnamefont
  {M.}~\bibnamefont {Minderer}}, \bibinfo {author} {\bibfnamefont
  {G.}~\bibnamefont {Heigold}}, \bibinfo {author} {\bibfnamefont
  {S.}~\bibnamefont {Gelly}}, \bibinfo {author} {\bibfnamefont
  {J.}~\bibnamefont {Uszkoreit}},\ and\ \bibinfo {author} {\bibfnamefont
  {N.}~\bibnamefont {Houlsby}},\ }\href@noop {} {\enquote {\bibinfo {title} {An
  image is worth 16x16 words: Transformers for image recognition at scale},}\ }
  (\bibinfo {year} {2021}),\ \Eprint {https://arxiv.org/abs/2010.11929}
  {arXiv:2010.11929 [cs.CV]} \BibitemShut {NoStop}%
\bibitem [{\citenamefont {Bradski}(2000)}]{opencv_library}%
  \BibitemOpen
  \bibfield  {author} {\bibinfo {author} {\bibfnamefont {G.}~\bibnamefont
  {Bradski}},\ }\bibfield  {title} {\enquote {\bibinfo {title} {{The OpenCV
  Library}},}\ }\href@noop {} {\bibfield  {journal} {\bibinfo  {journal} {Dr.
  Dobb's Journal of Software Tools}\ } (\bibinfo {year} {2000})}\BibitemShut
  {NoStop}%
\bibitem [{spm(2023)}]{spmu}%
  \BibitemOpen
  \href@noop {} {\enquote {\bibinfo {title} {Spmutil},}\ } (\bibinfo {year}
  {2023}),\ \bibinfo {note}
  {\url{https://github.com/DIAOZHUO/SPMUtil}}\BibitemShut {NoStop}%
\bibitem [{\citenamefont {Paszke}\ \emph {et~al.}(2019)\citenamefont {Paszke},
  \citenamefont {Gross}, \citenamefont {Massa}, \citenamefont {Lerer},
  \citenamefont {Bradbury}, \citenamefont {Chanan}, \citenamefont {Killeen},
  \citenamefont {Lin}, \citenamefont {Gimelshein}, \citenamefont {Antiga},
  \citenamefont {Desmaison}, \citenamefont {Kopf}, \citenamefont {Yang},
  \citenamefont {DeVito}, \citenamefont {Raison}, \citenamefont {Tejani},
  \citenamefont {Chilamkurthy}, \citenamefont {Steiner}, \citenamefont {Fang},
  \citenamefont {Bai},\ and\ \citenamefont {Chintala}}]{pytorch}%
  \BibitemOpen
  \bibfield  {author} {\bibinfo {author} {\bibfnamefont {A.}~\bibnamefont
  {Paszke}}, \bibinfo {author} {\bibfnamefont {S.}~\bibnamefont {Gross}},
  \bibinfo {author} {\bibfnamefont {F.}~\bibnamefont {Massa}}, \bibinfo
  {author} {\bibfnamefont {A.}~\bibnamefont {Lerer}}, \bibinfo {author}
  {\bibfnamefont {J.}~\bibnamefont {Bradbury}}, \bibinfo {author}
  {\bibfnamefont {G.}~\bibnamefont {Chanan}}, \bibinfo {author} {\bibfnamefont
  {T.}~\bibnamefont {Killeen}}, \bibinfo {author} {\bibfnamefont
  {Z.}~\bibnamefont {Lin}}, \bibinfo {author} {\bibfnamefont {N.}~\bibnamefont
  {Gimelshein}}, \bibinfo {author} {\bibfnamefont {L.}~\bibnamefont {Antiga}},
  \bibinfo {author} {\bibfnamefont {A.}~\bibnamefont {Desmaison}}, \bibinfo
  {author} {\bibfnamefont {A.}~\bibnamefont {Kopf}}, \bibinfo {author}
  {\bibfnamefont {E.}~\bibnamefont {Yang}}, \bibinfo {author} {\bibfnamefont
  {Z.}~\bibnamefont {DeVito}}, \bibinfo {author} {\bibfnamefont
  {M.}~\bibnamefont {Raison}}, \bibinfo {author} {\bibfnamefont
  {A.}~\bibnamefont {Tejani}}, \bibinfo {author} {\bibfnamefont
  {S.}~\bibnamefont {Chilamkurthy}}, \bibinfo {author} {\bibfnamefont
  {B.}~\bibnamefont {Steiner}}, \bibinfo {author} {\bibfnamefont
  {L.}~\bibnamefont {Fang}}, \bibinfo {author} {\bibfnamefont {J.}~\bibnamefont
  {Bai}},\ and\ \bibinfo {author} {\bibfnamefont {S.}~\bibnamefont
  {Chintala}},\ }\href@noop {} {\enquote {\bibinfo {title} {Pytorch: An
  imperative style, high-performance deep learning library},}\ } (\bibinfo
  {year} {2019})\BibitemShut {NoStop}%
\bibitem [{\citenamefont {Jocher}, \citenamefont {Chaurasia},\ and\
  \citenamefont {Qiu}(2023)}]{yolov8}%
  \BibitemOpen
  \bibfield  {author} {\bibinfo {author} {\bibfnamefont {G.}~\bibnamefont
  {Jocher}}, \bibinfo {author} {\bibfnamefont {A.}~\bibnamefont {Chaurasia}},\
  and\ \bibinfo {author} {\bibfnamefont {J.}~\bibnamefont {Qiu}},\ }\href@noop
  {} {\enquote {\bibinfo {title} {Ultralytics yolov8},}\ } (\bibinfo {year}
  {2023})\BibitemShut {NoStop}%
\bibitem [{\citenamefont {Buslaev}\ \emph {et~al.}(2020)\citenamefont
  {Buslaev}, \citenamefont {Iglovikov}, \citenamefont {Khvedchenya},
  \citenamefont {Parinov}, \citenamefont {Druzhinin},\ and\ \citenamefont
  {Kalinin}}]{albumentations}%
  \BibitemOpen
  \bibfield  {author} {\bibinfo {author} {\bibfnamefont {A.}~\bibnamefont
  {Buslaev}}, \bibinfo {author} {\bibfnamefont {V.~I.}\ \bibnamefont
  {Iglovikov}}, \bibinfo {author} {\bibfnamefont {E.}~\bibnamefont
  {Khvedchenya}}, \bibinfo {author} {\bibfnamefont {A.}~\bibnamefont
  {Parinov}}, \bibinfo {author} {\bibfnamefont {M.}~\bibnamefont {Druzhinin}},\
  and\ \bibinfo {author} {\bibfnamefont {A.~A.}\ \bibnamefont {Kalinin}},\
  }\bibfield  {title} {\enquote {\bibinfo {title} {Albumentations: Fast and
  flexible image augmentations},}\ }\href
  {https://doi.org/10.3390/info11020125} {\bibfield  {journal} {\bibinfo
  {journal} {Information}\ }\textbf {\bibinfo {volume} {11}} (\bibinfo {year}
  {2020}),\ 10.3390/info11020125}\BibitemShut {NoStop}%
\bibitem [{\citenamefont {Loshchilov}\ and\ \citenamefont
  {Hutter}(2019)}]{adamw}%
  \BibitemOpen
  \bibfield  {author} {\bibinfo {author} {\bibfnamefont {I.}~\bibnamefont
  {Loshchilov}}\ and\ \bibinfo {author} {\bibfnamefont {F.}~\bibnamefont
  {Hutter}},\ }\href {https://openreview.net/forum?id=Bkg6RiCqY7} {\enquote
  {\bibinfo {title} {Decoupled weight decay regularization},}\ } (\bibinfo
  {year} {2019})\BibitemShut {NoStop}%
\end{thebibliography}%
%\input ai.bbl

\end{document}